#\documentclass[linenumbers,twocolumn]{aastex631}
\documentclass[twocolumn]{aastex631}

\usepackage{amsmath}

\begin{document}

\title{
X-ray Polarization of the High-Synchrotron-Peak BL Lacertae Object 1ES~1959+650 during Intermediate and High X-ray Flux States}

\correspondingauthor{Luigi Pacciani}
\email{luigi.pacciani@inaf.it}

\author[0000-0001-6897-5996]{Luigi Pacciani}
\affiliation{INAF, Istituto di Astrofisica e Planetologia Spaziali, Via Fosso del Cavaliere, 100 - I-00133 Rome, Italy}

\author[0000-0001-5717-3736]{Dawoon E. Kim}
\affiliation{INAF, Istituto di Astrofisica e Planetologia Spaziali, Via Fosso del Cavaliere, 100 - I-00133 Rome, Italy}
\affiliation{Dipartimento di Fisica, Universit\`a degli Studi di Roma “La Sapienza”, Piazzale Aldo Moro 5, 00185 Roma, Italy}
\affiliation{Dipartimento di Fisica, Universit\`a degli Studi di Roma “Tor Vergata”, Via della Ricerca Scientifica 1, 00133 Roma, Italy}

\author[0000-0001-9815-9092]{Riccardo Middei}
\affiliation{INAF, Istituto di Astrofisica e Planetologia Spaziali, Via Fosso del Cavaliere, 100 - I-00133 Rome, Italy}

\author[0000-0002-6492-1293]{Herman L. Marshall}
\affiliation{MIT Kavli Institute for Astrophysics and Space Research, Massachusetts Institute of Technology, 77 Massachusetts Avenue, Cambridge, MA 02139, USA}

\author[0000-0001-7396-3332]{Alan P. Marscher}
\affiliation{Institute for Astrophysical Research, Boston University, 725 Commonwealth Avenue, Boston, MA 02215, USA}

\author[0000-0001-9200-4006]{Ioannis Liodakis}
\affiliation{NASA Marshall Space Flight Center, Huntsville, AL 35812, USA}
\affiliation{Institute of Astrophysics, Foundation for Research and Technology-Hellas, GR-71110 Heraklion, Greece}

\author[0000-0002-3777-6182]{Iv\'{a}n Agudo}
\affiliation{Instituto de Astrof\'{i}sica de Andaluc\'{i}a, IAA-CSIC, Glorieta de la Astronom\'{i}a s/n, 18008 Granada, Spain}

\author[0000-0001-6158-1708]{Svetlana G. Jorstad}
\affiliation{Institute for Astrophysical Research, Boston University, 725 Commonwealth Avenue, Boston, MA 02215, USA}
\affiliation{Saint Petersburg State University, 7/9 Universitetskaya nab., St. Petersburg, 199034 Russia}

\author[0000-0002-0983-0049]{Juri Poutanen}
\affiliation{Department of Physics and Astronomy, 20014 University of Turku, Finland}

\author[0000-0002-1853-863X]{Manel Errando}
\affiliation{Physics Department and McDonnell Center for the Space Sciences, Washington University in St, Louis, MO, 63130, USA}

% ...

\begin{abstract}
We report the \textit{Imaging X-ray Polarimetry Explorer} (IXPE) polarimetric
and simultaneous multiwavelength observations of the
high-energy-peaked BL Lacertae (HBL) object 1ES~1959+650, performed in 2022 October and 2023 August.
In 2022 October IXPE measured an average polarization degree  $\Pi_{\rm X}=9.4\;\!\%\pm 1.6\;\!\%$
and an electric-vector position angle $\psi_{\rm X}=53\degr\pm 5\degr$.
The polarized X-ray emission can be decomposed into a constant component, plus a rotating component,
with rotation velocity $\omega_{\rm EVPA}=(-117\;\!\pm\;\!12)$\;\!deg~d$^{-1}$. In 2023 August,
during a period of pronounced activity of the source,
IXPE measured an average $\Pi_{\rm X}=12.4\;\!\%\pm0.7\;\!\%$ and $\psi_{\rm X}=20\degr\pm2\degr$,
with evidence ($\sim$0.4$\;\!\%$ chance probability) for 
a rapidly rotating component with  $\omega_{\rm EVPA}=1864\;\!\pm\;\!34$deg~d$^{-1}$.
These findings suggest the presence of a helical magnetic field in the jet of 1ES~1959+650
or stochastic processes governing the field in turbulent plasma.
Our multiwavelength campaigns from radio to X-ray reveal variability in both polarization and flux
from optical to X-rays. We interpret the results in terms of a relatively slowly varying component  
dominating the radio and optical emission, while rapidly variable polarized components dominate
the X-ray and provide minor contribution at optical wavelengths.
The radio and optical data indicate that on parsec scales the magnetic field is primarily orthogonal
to the jet direction. On the contrary, X-ray measurements show a magnetic field almost aligned
with the parsec jet direction. Confronting with other IXPE observations, we guess that the magnetic
field of HBLs on sub-pc scale should be rather unstable, often changing its direction with respect
to the VLBA jet.
\end{abstract}


\section{Introduction} \label{sec:intro}
Blazars are active galactic nuclei (AGN) with a relativistic jet of plasma pointing within several degrees of the line of sight \citep[e.g.,][]{BlandfordKonigl1979}. Relativistic motion (with bulk Lorentz factors $\Gamma\sim 10$)
beams the radiation \citep{UrryPadovani1995}, causing these objects to be the most luminous persistent extragalactic
sources at wavelengths from radio to TeV $\gamma$-ray. The non-thermal emission from the blazar jet can be extremely variable, on timescales as short as several minutes \citep{Aharonian2007,Ackermann2016}.

BL Lacertae objects are a class of blazars that exhibit little  or no thermal emission, in lines or in continuum, in their near-IR, optical, and UV spectra. Hence, their spectral energy distribution (SED) is entirely dominated by non-thermal processes. High-energy-peaked BL Lac objects (HBLs) are a sub-class with the SED of their synchrotron emission peaking at the X-ray energies.

The target of this investigation, 1ES~1959+650, is an HBL at redshift $z=0.047$
\citep{Perlman1996}. It is among the first blazars detected at TeV energies (e.g., \citealt{Aharonian2003}), and it has been the target of several multifrequency campaigns (e.g., \citealt{Krawczynski2004,Tagliaferri2008,Aliu2014,2020A&A...638A..14M}).
Very Long Baseline Array (VLBA) images obtained between 2005 and 2009 (and similarly for Mrk~501 and Mrk~421) at 22~GHz \citep{2010ApJ...723.1150P} revealed a characteristic polarization structure of the jet: the electric vector position angle (EVPA) was parallel to the jet axis (position angle $\sim150^\circ$, measured north through east) in the ``core'' (upstream end on the images) of the jet, while on the edges it was closer to orthogonal. This pattern can be interpreted in terms of a structured (spine-sheath) jet \citep{2010ApJ...723.1150P}. Later, higher-resolution VLBA images at 43 GHz measured the jet direction in the most compact region to be $128\degr\pm13\degr$ \citep{Weaver2022}.

The EVPA of the optical polarization of 1ES~1959+650, observed between 2009 and 2011, varied about a stable value of $\sim150^\circ$ \citep[i.e., parallel to the compact jet;][]{2013ApJS..206...11S}. 
The authors speculated that two components
are responsible for the optical emission, with one nearly stable component with EVPA $\sim$ $150^\circ$  associated with the overall jet emission, and a time varying component, perhaps originating in shocks within the jet.

The light curves of 1ES~1959+650 between 2016 and 2017 contain a number of flares
from radio to TeV energies. These flux variations and the SED can be explained by a model with two electron populations: one with a synchrotron component peaking at radio-IR frequencies and another, highly variable component peaking in the X-ray range, with the optical emission corresponding to the superposition of both emitting components \citep{patel2018,2020A&A...638A..14M,2020A&A...640A.132M}; but see also \cite{2018ApJS..238...13K} and \cite{2021ApJ...918...67C} for different interpretations.

The \textit{Imaging X-ray Polarimetry Explorer} \citep[IXPE;][]{Weisskopf2022} observed 1ES~1959+650 twice during the first semester of scientific operations. Polarization was  detected during the first pointing (2022 May) with the polarization degree (PD) $\Pi_{\rm X}=8.0\;\!\%\pm2.3\;\!\%$, although only an upper limit ($\Pi_X<5.1\% ,\;\! 99\;\!\%\;\!$ confidence level, c.l.) was obtained from the second observation (2022 June) of the source \citep{Errando2024}. During the second campaign, optical polarization at a level of $\Pi_{\rm O}\sim$5\;\!\% was measured.\\
IXPE also observed the HBLs Mrk~421, and Mrk~501 during the first semester of scientific operations, detecting linear polarization in all of them \citep{2022Natur.611..677L,2022ApJ...938L...7D}.
The Mrk~501 observations found an almost constant X-ray PD and EVPA. In contrast with the second campaign of 1ES~1959+650,
the X-ray PD was measured to be a factor of 
2--2.5 times higher than at optical wavelengths, and
the X-ray EVPA was found to be close to the  optical and radio values, and also close to the direction of the jet axis. This is compatible with a model in which particles are accelerated at a shock front, after which they lose energy, leading to an energy-stratified emission region \citep{2022Natur.611..677L}.\\
The first IXPE observation of Mrk~421 \citep{2022ApJ...938L...7D} revealed a similar phenomenology as in the case
of Mrk~501, but without alignment of the X-ray and optical EVPAs. Two subsequent IXPE
observations of Mrk~421 found a rotation of the EVPA \citep{DiGesu2023}. At the time of the X-ray EVPA rotation, the radio and optical PDs were lower than the X-ray values, and did not show any significant variation. 
During the second year of IXPE observations, four additional HBL objects were observed: 1ES~0229+200 \citep{Ehlert2023}, PG~1553+113 \citep{Middei2023-II}, PKS~2155-304 \citep{Kouch2024}. These displayed behavior mostly consistent with that of Mrk~501. However, in the case of PG~1553+113, during the IXPE observation an optical EVPA rotation was detected without a radio or X-ray counterpart.\\

Here we report the results of two additional sets of IXPE and  multi-wavelength observations of 1ES~1959+650, one in 2022 late October and another in 2023 August. The latter campaign was triggered by an X-ray outburst \citep{Kapanadze2023}.
\section{Log of X-ray observations}
 The two sets of IXPE observations took place
(1) between 2022-10-28 at 06:02 UT and 2022-10-31 at 12:16 UT (2022 October campaign); and
(2) between 2023-08-14 at 01:09 UT and 2023-08-19 at 08:50 UT (2023 August campaign).
During the 2022 October campaign, XMM-Newton observed the source starting on 2022-10-28 17:11 UT for 17.0\;\!ks. MOS1 was operated in PrimePartialW2 mode with Thick filter, MOS2 in FastUncompressed mode with Medium filter, and PN in timing mode with Thick filter.\\
NuSTAR observed 1ES~1959+650 starting on 2022-10-31 at 00:16:09 UT for 17.8\;\!ks, and Swift observed several times with the XRT, always operating in Windowed-Timing mode, and with the UVOT using all 6 filters.

During the 2023 August campaign, Swift observed the source with one-day cadence with the XRT always in Windowed-Timing mode, and the UVOT using all 6 filters.\\
A summary of the X-ray observations is given in Table~\ref{table:logx}.
\begin{table}
\footnotesize
    \centering
    \begin{tabular}{c c c} \\ \hline \hline
    X-ray              &  Observation start         & exposure   \\
    observatory        &    (UT)                 &  (ks)      \\
     IXPE              &    2022-10-28 06:02                         & 281.6\\
     XMM-Newton        &    2022-10-28 17:11                         & 17.0\\
     NuSTAR            &    2022-10-31 00:16                         & 17.8 \\
     Swift-xrt         &    2022-10-29 09:39                         & 0.9 \\
                       &    2022-10-30 03:08                         & 0.7 \\
                       &    2022-10-31 01:16                         & 1.7  \\
                       \\
     IXPE              &   2023-08-14 01:09                           & 459.7\\
     Swift-xrt         &   2023-08-14 17:52                           & 1.7 \\ 
                       &   2023-08-15 19:12                           & 1.7 \\ 
                       &   2023-08-16 01:26                           & 1.0 \\ 
                       &   2023-08-16 23:58                           & 0.8 \\ 
                       &   2023-08-18 20:05                           & 1.1 \\
                       &   2023-08-19 00:52                           & 0.8 \\   \hline
\end{tabular}
 \caption{Log of X-ray observations}
 \label{table:logx}
 \end{table}
\section{Data Analysis}
\subsection{IXPE data} \label{sec:ixpe}
We used IXPE level 2 data, with photon-by-photon information on time of arrival, position, energy, and the $Q$ and $U$ Stokes parameters.
We analysed data using the publicly available \textsc{ixpeobssim} software version 30.5.0 \citep{2022SoftX..1901194B}.
We extracted the source data from a circular region with a 1.2\arcmin\ radius centered on the source position, while we extracted background data from an annular region with  2.5\arcmin\ and 3.5\arcmin\ inner and outer radii, respectively.
We used the \texttt{xpbin} procedure to obtain the polarization cube (PCUBE) and the Stokes parameter spectra ($I$, $Q$, and $U$); we subtracted background by applying the procedure proposed in \cite[][]{2022SoftX..1901194B}. 
We generated source and background PCUBEs for each detector unit. Detector-by-detector spectra were produced for $I$, $Q$, $U$ Stokes parameters using the PHAI, PHAQ, and PHAU methods in \texttt{xpbin} and applying calibration database
version CALDB 20221020 and a background-to-source BACKSCALE ratio of 0.05. The weighted analysis algorithm proposed in \citet{2022AJ....163..170D} was used with 075 response matrices for the spectral analysis.
A light curve was produced starting with a bin size of 6\;\!s. The time bins not contained in the GTIs were rejected, while the others were grouped in order to obtain the final bin size.

\subsection{Swift-XRT data}
We reduced Swift-XRT \citep{Burrows2005} data using \texttt{xrtpipeline} version 0.13.4, included in the \textsc{HEASOFT} v6.25 package,
and using the most recent available calibration
files. Events with grade 0–2 were selected for WT data, and with grade 0--12 for photon
counting (PC) mode. We used \texttt{xrtmkarf} to create the Ancillary Response files.
Several Swift-XRT observations span more than a satellite orbit. In order to give a detailed light curve, we subdivided 
these observations on an orbit-by-orbit basis. The obtained source counting rate reported in the Swift-XRT light curve
was then corrected for vignetting and for source signal lost in dead strips in WT mode.

\subsection{NuSTAR data}
We made use of standard level 2 event files generated
by the NuSTAR SOC and available from HEASARC archive.
We reduced and analyzed NuSTAR data using the \textsc{NuSTARDAS1} Data Analysis Software in the
\textsc{HEASoft V6.29} package, adopting CALDB version 20211020 calibration files.
We used the nuproducts tool to extract high-level science products for the source
in the 3$-$20 keV energy range.
The source events were selected from a circular extraction region of 30\arcsec\ radius,
while the background was computed in an annulus centered on the source
with inner and outer radii of 50\arcsec\ and 100\arcsec\, respectively.
\subsection{XMM-Newton data}
We employed the XMM-Newton Science Analysis System, version 18.0.0, to process the XMM-Newton data.
Source events were extracted from a circular region with 30\arcsec\ radius centered on the source, while background events
were obtained from a circular region  with 40\arcsec\ radius, offset from the source.
Spectra were re-binned to have at least 30 counts for each spectral bin.
\subsection{Swift/UVOT data}
We analyzed Swift-UVOT \citep{Roming2005} data using the \textsc{HEASOFT} \texttt{uvotimsum} and \texttt{uvotsource} procedures. Source flux was extracted through aperture photometry from 
a 5\arcsec\ circular region, and the
UV magnitudes were corrected with values from \citet{Pei1992}. Fluxes were derived from magnitudes according to \citet{Poole2008}. 

\subsection{Radio/Optical data}
During the IXPE observations, 
several ground-based telescopes at radio 
(10$-$225~GHz) and optical (BVRI bands) frequencies provided total flux and polarization measurements. These included the Effelsberg 100-m telescope, the Submillimeter Array (SMA), Calar Alto Observatory, the LX-200 telescope operated by Saint Petersburg University, the Nordic Optical Telescope (NOT), the 1.8~m Perkins telescope owned by Boston University (Perkins Telescope Observatory, PTO), the T90 telescope at the Sierra Nevada Observatory (SNO), and the RoboPol 1.3~m telescope at Skinakas Observatory.
Logs of optical and radio observations are given in Table~\ref{table:logo} and Table~\ref{table:logr}, respectively.\\
\begin{table}
\footnotesize
    \centering
    \begin{tabular}{c c c} \\ \hline \hline
   Observatory  & filter   & date \\
                &          & (UT) \\
   NOT   &  B$^p$,V$^p$,R$^p$I$^p$  & 2022-10-30 20:09  \\ 
         &  B$^p$,V$^p$,R$^p$I$^p$  & 2022-10-31 20:52 \\
   Calar Alto  &  R$^p$             & 2022-10-29 20:25 ($^M$) \\
               & R$^p$             & 2022-10-29 20:26 ($^M$)  \\       
   SNO         & R                 & 2022-10-27 22:38  ($^M$) \\
               & R$^p$             & 2022-10-28 22:30  ($^M$) \\
   S. Petersburg & R$^p$,I         & 2022-10-30 16:24 \\
                 & R$^p$,I         & 2022-10-31 18:37 \\

   Perkins       & R$^p$                    & 2022-10-25 02:52 ($^M$)\\
                & R$^p$                     & 2022-10-26 02:52  \\
                 & V$^p$,I$^p$,B$^p$,R$^p$  & 2022-10-27 03:42 \\
                 & V$^p$,I$^p$,B$^p$,R$^p$  & 2022-10-28 03:12 \\
                 & V$^p$,I$^p$,B$^p$,R$^p$  & 2022-10-29 02:36 \\
                & V$^p$,I$^p$,B$^p$,R$^p$  & 2022-10-29 04:22 \\
                & V$^p$,I$^p$,B$^p$,R$^p$  & 2022-10-30 02:34 \\
                & V$^p$,I$^p$,B$^p$,R$^p$  & 2022-10-30 04:25 \\
                & V$^p$,I$^p$,B$^p$,R$^p$  & 2022-10-31 02:58  \\
                & V$^p$,I$^p$,B$^p$,R$^p$  & 2022-10-31 04:52 \\
                & V$^p$,I$^p$,B$^p$,R$^p$  & 2022-11-01 03:07 \\
                \\
   NOT   &  B$^p$,V$^p$,R$^p$,I$^p$  &  2023-08-14 02:52 \\
         & B$^p$,V$^p$,R$^p$,I$^p$  &   2023-08-15 04:04 \\
         & B$^p$,V$^p$,R$^p$,I$^p$  &  2023-08-17 03:07  \\
         & B$^p$,V$^p$,R$^p$,I$^p$  &  2023-08-19 02:24  \\
         & B$^p$,V$^p$,R$^p$,I$^p$  &  2023-08-20 07:12 \\            
         & B$^p$,V$^p$,R$^p$,I$^p$  &   2023-08-21 05:01 \\       
Calar Alto & R$^p$  &  2023-08-16 22:33 ($^M$)\\
           & R$^p$  & 2023-08-18 02:09 ($^M$)\\
SNO        &  R     &  2023-08-14 23:45\\
           &  R     &  2023-08-18 00:43 ($^M$)\\
           & R$^P$  & 2023-08-19 00:08 ($^M$)\\ 
           & R$^p$  & 2023-08-20 00:28  ($^M$)\\ 
           & R      & 2023-08-22 01:32  \\
           & R      & 2023-08-24 22:04 \\
           & R      & 2023-08-27 22:01 \\
S. Petersburg  & B,V,R$^p$,I  &  2023-08-14 22:19 ($^M$)\\ 
               & B,V,R$^p$,I  &  2023-08-15 22:04 ($^M$)\\ 
               & B,V,R$^p$,I  &  2023-08-17 21:43 ($^M$)\\ 
               & B,V,R$^p$,I  & 2023-08-20 21:41  ($^M$)\\ 
Skinakas        & R$^p$        & 2023-08-03 19:26 \\
                & R$^p$       & 2023-08-04 19:44  \\
                & R$^p$        & 2023-08-05 19:59 \\
                & R$^p$        &  2023-08-06 22:56 \\              
                & R$^p$        & 2023-08-16 18:16 \\
                & R$^p$        & 2023-08-18 17:59 \\
                & R$^p$        & 2023-08-20 17:47 \\
                & R$^p$        & 2023-08-24 19:40 \\
                & R$^p$        & 2023-08-26 22:17 \\             
                & R$^p$        & 2023-08-28 18:53 \\
\end{tabular}
 \caption{Log of optical observations. The letter $p$ on top of the filter flags a polarimetric measurement; The letter $M$ after the observing date flags multiple observations during the same night. Swift-uvot observations, performed with all six filters, were performed simultaneously to all the Swift-xrt observations reported in Table~\ref{table:logx}.}
 \label{table:logo}
\end{table}

\begin{table}
\footnotesize
\centering
    \begin{tabular}{c c c} \\ \hline \hline
   Observatory  & Frequency   & date \\
                &   (GHz)       & (UT) \\
SMA   & 225 & 2022-11-10 07:07:24 \\ 
\\
SMA   & 225 & 2023-08-06 10:19 \\
      & 225 &  2023-08-16 06:57 \\
      & 225 & 2023-08-17 06:59  \\
      & 225 & 2023-08-18 07:55 \\
      & 225 & 2023-08-19 06:48 \\
      & 225 & 2023-08-20 07:26 \\
Effelsberg & 10.45, 17 & 2023-07-29 23:16 \\
           & 10.45, 17 & 2023-08-10 23:31 \\
           & 10.45, 17 & 2023-08-15 19:12 \\
          & 10.45, 17 & 2023-08-17 19:40  \\
          & 10.45, 17 & 2023-08-21 01:12 \\
\end{tabular}
 \caption{Log of Radio observations.}
 \label{table:logr}
\end{table}

The Effelsberg 100-m telescope observations were taken at 10.45~GHz and 17~GHz in July and 2023 August, as part of the QUIVER (monitoring the Stokes $Q$, $U$, $I$, and $V$ emission of AGN jets in Radio) monitoring program \citep{Kraus2003,Myserlis2018}. The SMA observations were obtained at 1.3~mm (225~GHz) on 2022 October 31, and several dates in 2023 August, within the SMAPOL (SMA Monitoring of AGN with POLarization) program. The observing setup involved orthogonally polarized receivers in full polarization mode. The polarized intensity, degree, and angle were derived from Stokes parameters $I$, $Q$, and $U$ and calibrated using the MIR package\footnote{\url{https://lweb.cfa.harvard.edu/~cqi/mircook.html}} \citep{Saul1995,Marrone2008,Primiani2016}.

The Calar Alto and Sierra Nevada observations were taken in R band using the CAFOS polarimeter and a set of polarized filters, respectively. Several observations were taken 
for several nights during each IXPE observation. For SNO we used the weighted average of all the observations taken within the same night. The NOT observations were taken in the BVRI bands during both IXPE pointings, using the Alhambra Faint Object Spectrograph and Camera (ALFOSC). The observations and data reduction are described in detail in \cite{Nilsson2018}. Additional photometric and polarimetric measurements in BVRI bands were obtained at the Perkins telescope (Flagstaff, Arizona) from  2022 October 27 to 2022 November 1 using the PRISM camera. A general description of polarimetric observations can be found in
\cite{Jorstad2010}. The Skinakas observations were taken in R-band during 2023 August using the 4-channel RoboPol polarimeter mounted on the 1.3~m telescope \citep{Panopoulou2015,Blinov2021}. Finally, the LX-200 telescope provided R- and I-band total intensity and R-band polarimetric observations on 2022 October 30  and 2022 October 31, and BVRI photometry and R-band polarimetry during the 2023 August observation. A description of observational procedures and data reduction can be found in \cite{Larionov2008}. More details on the IXPE-related multiwavelength observing strategy and data reduction can be found in \cite{2022Natur.611..677L,Middei2023,Peirson2023,Kouch2024}, and \cite{DiGesu2023}. The R-band polarization measurements were corrected for dilution of the polarization by unpolarized host-galaxy starlight by subtracting its contribution within the apertures used by different observatories, following \cite{Nilsson2007,Hovatta2016}. For the remaining optical bands we report the observed values.

\section{Results} \label{sec:results}

\subsection{Light Curves and Hardness Ratios}
\begin{figure*}[tb]
\centering
\begin{tabular}{ll}
\includegraphics[width=0.4\textwidth]{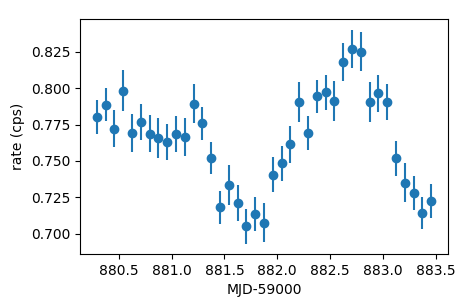}   &
\includegraphics[width=0.4\textwidth]{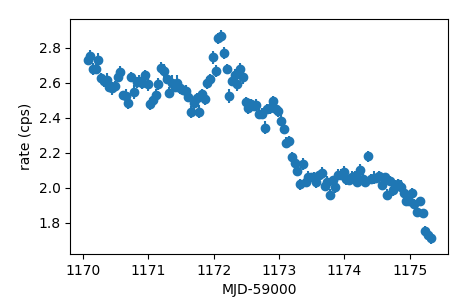} 

\end{tabular}
\caption{IXPE $2-8$~keV light curves. \textit{Left:} 2022 October campaign; \textit{right:} 2023 August campaign.}
\label{fig:ixpe_lc}
\end{figure*}

The Swift-XRT and Swift-UVOT light curves are presented 
in Figure \ref{fig:lc_xrt_uvot}, covering the two observing campaigns.
The first IXPE campaign was performed during a period of intermediate activity of the source, while the second was performed during a bright state.
The IXPE light curves for the two periods are displayed in Figure~\ref{fig:ixpe_lc}.

\begin{figure*}[tb]
\centering
\begin{tabular}{ll}
\includegraphics[width=0.4\textwidth]{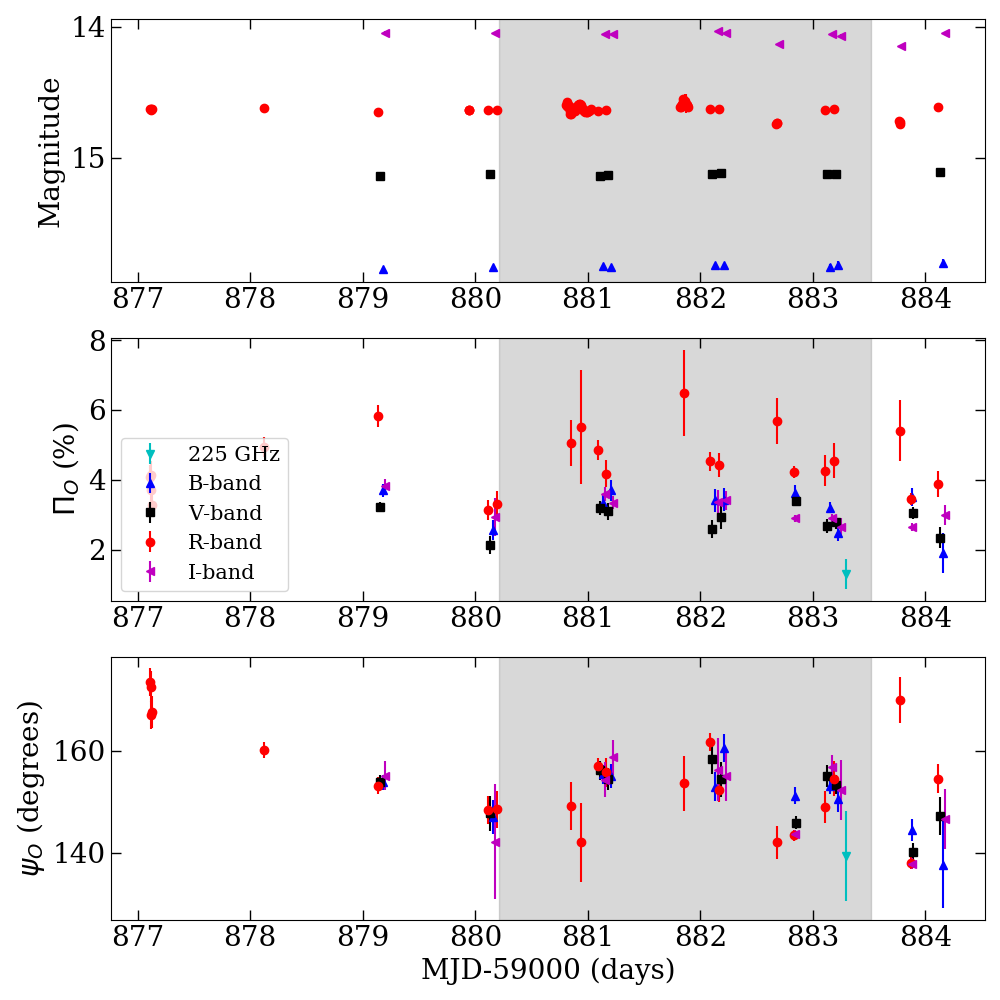}   & \includegraphics[width=0.4\textwidth]{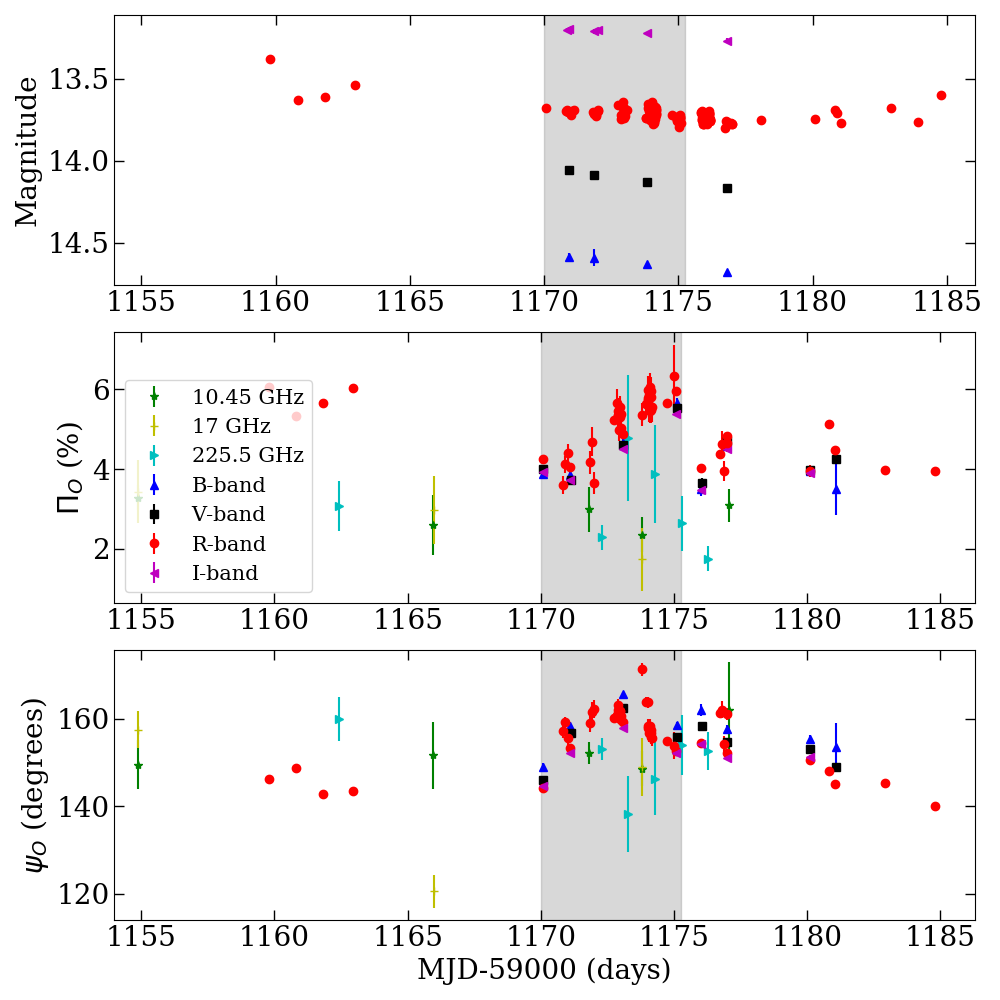} 
\end{tabular}
\caption{Radio and optical light curves and polarimetric measurements with ground-based telescopes. 
The 2022 October campaign in the left panel; 
the 2023 August campaign in the right panel. 
The panels from top to bottom show optical brightness, radio and optical PD, and radio and optical polarization angle. The duration of the IXPE observation is indicated by the grey shaded area. }
\label{fig:radio_opt_lc_pol}
\end{figure*}

The hardness ratio for the Swift-XRT data (defined as the ratio 
$({H-S})/({H+S})$ with $H$ and $S$ being the photon counts in the 
2--10~keV and 0.3--2~keV bands, respectively) is shown in Figure~\ref{fig:ixpe_lc};
\begin{figure*}[tb]
\centering
\includegraphics[width=1.1\textwidth]{swift_lcs_4plots_new.eps}
\caption{Swift light curves. \textit{Top to bottom:} UVOT V optical filter, UVOT UVW2 ultraviolet filter, XRT X-ray, and X-ray hardness ratio (evaluated adopting the 0.3--2 keV and the 2--10~keV bands). 
The IXPE observing periods are denoted with green areas.}
\label{fig:lc_xrt_uvot}
\end{figure*}
Radio and optical light curves for the two campaigns are displayed in Figure \ref{fig:radio_opt_lc_pol}.

\subsection{Polarimetry}
\subsubsection{X-ray polarimetry -- 2022 October campaign}
During the 2022 October observation, IXPE detected an average X-ray PD of $9.4\;\!\%\pm 1.6\;\!\%$ and an
EVPA of $53\degr\pm 5\degr$, obtained by integrating over the entire 3.3-day pointing and over the energy interval 
$2-8$\;\!keV.
For comparison, the minimum detectable polarization (corresponding to 99\% c.l.) for that period was 4.9\;\!\%.

\begin{figure*}[tb]
\centering
\begin{tabular}{ll}
\includegraphics[width=0.47\textwidth]{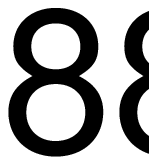} & \includegraphics[width=0.45\textwidth]{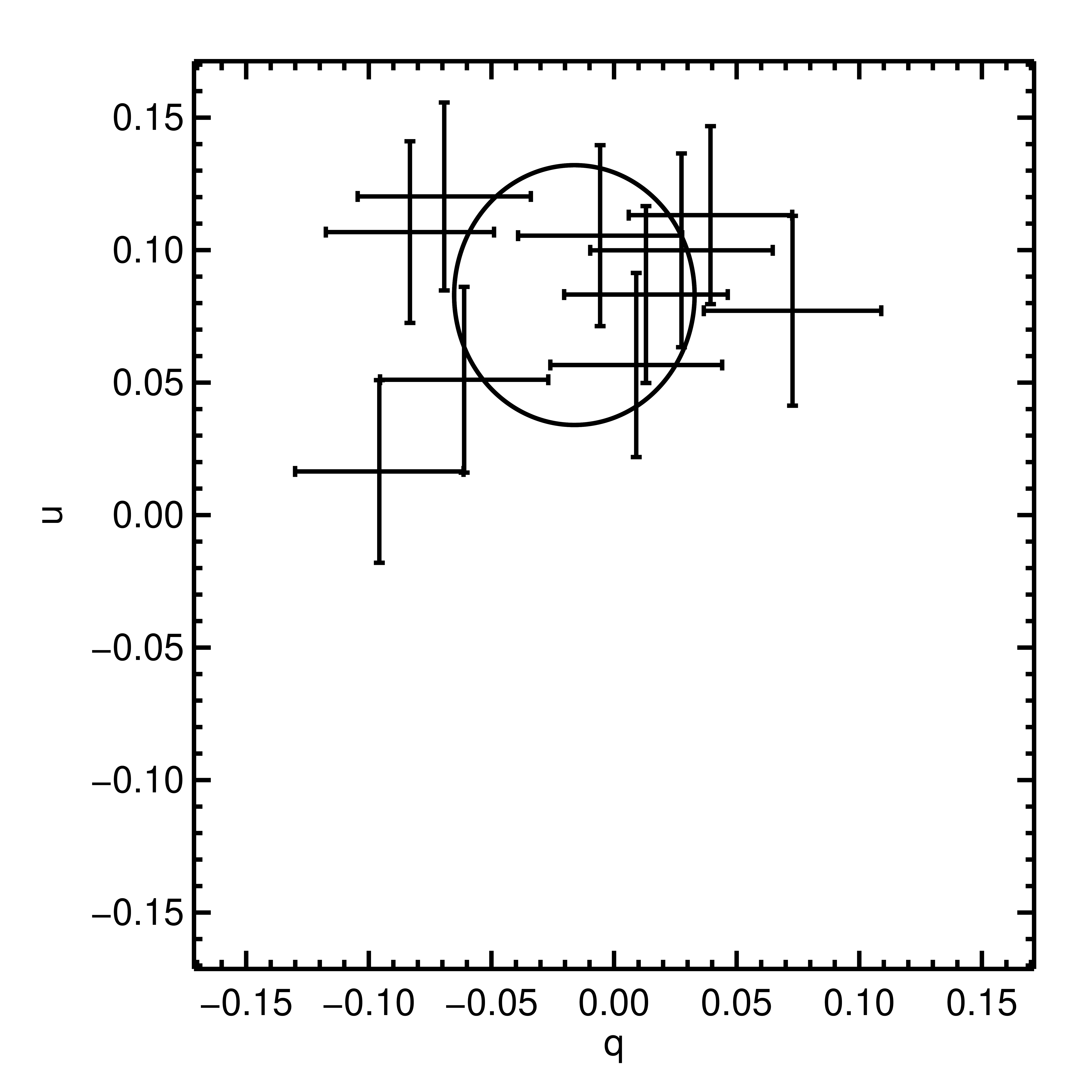} \\
\end{tabular}
\caption{Left panel: Polarimetric evolution for the 2022 October observation. The constant polarization model (dotted line) and the two-component model (continuous line) are superposed on the data points. Right panel: $q$ vs.\ $u$ plot for the 2022 October observation. The continuous line is the two-component model fitting function.}
\label{fig:qu_oct2022}
\end{figure*}

\begin{figure*}[tb]
\centering
\begin{tabular}{lll}
\includegraphics[width=0.32\textwidth]{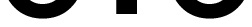} & \includegraphics[width=0.32\textwidth]{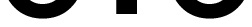} & \includegraphics[width=0.32\textwidth]{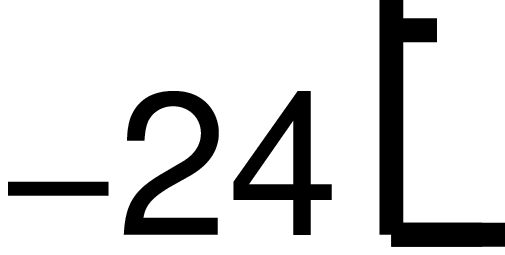} \\
\end{tabular}
\caption{Results of binned analysis fitting of the 2022 October polarimetric light curve. Fit was performed on the $q$ and $u$ Stokes light curves. Left panel: reduced $\chi^2$ for the constant polarization model as a function of the number of time bins, and associated probability that the fit function is true. Central panel: Reduced $\chi^2$ for the two-component polarization model as a function of the number of time bins, and associated probability that the fit function is true. Right panel: difference between the   $\chi^2$ for the two-component model and the $\chi^2$ for the constant component as a function of the number of time bins (details are in Appendix \ref{appendix:signifA}).}
\label{fig:const_vs_bicomp_oct2022}
\end{figure*}

The X-ray polarimetric evolution is shown in Figure~\ref{fig:qu_oct2022} in $q$,$u$ space, together with the constant polarization model. The result of  fitting with a constant polarization model in $q$,$u$ space is reported in the left panel of Figure~\ref{fig:const_vs_bicomp_oct2022} as a function of the number of time bins of the polarimetric light curve; the probability that the constant polarization model is correct is reported in the same figure. For the majority of chosen numbers of bins (except in four cases), the probability that the hypothesis of constant polarization is true is below 5\%. Therefore, we evaluated other models to describe the polarimetric data.

We investigated the possibility of global EVPA rotation during the observation, e.g., the case where EVPA rotation applies to the entire time span of observed X-ray emission. We  applied the unbinned likelihood method reported in \cite{DiGesu2023}, and we obtained an average global EVPA rotation rate $\omega_{\rm EVPA}\ =\ 
5 \pm 4$~deg~d$^{-1}$  (error at the 68\;\!\% c.l.); the rotation rate over the entire period is compatible with zero at 90\% c.l.\\
The data reported in  Figure~\ref{fig:qu_oct2022}, the IXPE observation of X-ray EVPA rotation in Mrk~421 \citep{DiGesu2023}, and the claim of two components contributing to the optical polarization and to the broadband SED of 1ES~1959+650, motivated us 
to test two nested hypotheses:\\ (a) The observed polarized X-rays have an EVPA that is steady, $\frac{{\rm d}\psi}{{\rm d}t}=0$ (steady polarization case); and\\
(b) The emission has
two superposed components: one with steady polarization and the other with variable EVPA with constant rotation rate $\omega_{\rm EVPA}$ (two-component model). The two-component model differs from the global EVPA rotation model by adopting the hypothesis that the EVPA rotation is not a global phenomenon of the entire X-ray emission of the blazar; rather, it applies to only a fraction of the observed emission.
The outcome of the tests of the two hypotheses is
given in Table \ref{tab:cash}, where we report the results of the minimization of the unbinned log-likelihood estimators explained in Appendix \ref{appendix:cash}.
\begin{table*}
    \centering
     \begin{tabular}{cccccccccc} \cline{1-2} \cline{5-6} \cline{9-10} 
    \noalign{\vskip\doublerulesep
         \vskip-\arrayrulewidth} 
     \cline{1-2} \cline{5-6}  \cline{9-10}
    \multicolumn{2}{c}{steady polarization model} & & & \multicolumn{2}{c}{two-component model}  & & & \multicolumn{2}{c}{Flux-correlated two-component model} \\ 
      C   & $-62.2$ & & & C & $-81.8$ &&& C & -85.3 \\
      $\Pi\ (\%) $   &  $9.4\pm 1.6$ & & & $R_1\Pi_1\ (\%)$ & $8.5^{+1.6}_{-1.9}$  & & & $\Pi_1\ (\%)$ & $10\pm2$ \\ 
      $\Psi\ ({\rm deg})$    & $53\pm 5$ & & & $\Psi_1\ ({\rm deg})$ & $51\pm6$  & & & $\Psi_1\ ({\rm deg})$ & $50\pm6$ \\
      \cline{1-2}   
      & & & & $R_2\Pi_2\ (\%)$ & $4.9\pm 1.7$  & & & $\Pi_2\ (\%)$ & $62^{+0.21}_{-0.48}$ \\
      & & & & $\Psi_2(t=0)\ ({\rm deg})$ & $31\pm21$  & & & $\Psi_2(t=0)\ ({\rm deg})$ & $25\pm21$ \\
      & & & & $\omega_2\ ({\rm deg\ d}^{-1})$ & $-117\pm 12$ & & & $\omega_2\ ({\rm deg\ d}^{-1})$ & $-117\pm 12$\\
    \cline{5-6} 
     & & & & & & & & r$_{1}$ (cts s$^{-1}$) & $0.71^{+0}_{-0.17}$ \\
  \cline{9-10} \\
    \end{tabular}
    \caption{Parameters and log-likelihood minima for the steady polarization model, for the two-component model, and for the flux-correlated two-component model
    for the 2022 October observation of 1ES~1959+650 with IXPE.
    $C$ is the log-likelihood minimum, $\Pi$ is the PD, and $\Psi$ is the EVPA for the steady polarization model. The two-component model has 5 parameters: $R_1\Pi_1$ is the product of the relative flux with the PD of component 1, and $R_2\Pi_2$ is the same for component 2. $\Psi_1$ is the EVPA of component 1, $\Psi_2(t=0)$ is the EVPA of rotating component 2 at the beginning of the observation, and $\omega_2$ is the angular velocity of the EVPA of the second component (see text).
  The flux-correlated two-component model has 6 parameters: $\Pi_1$ is the PD of component 1, and $\Pi_2$ is the same for component 2. $\Psi_1$ is the EVPA of component 1, $\Psi_2(t=0)$ is the EVPA of rotating component 2 at the beginning of the observation, $\omega_2$ is the angular velocity of the EVPA of the second component, and r$_1$ is the count rate of the steady component.}   
    \label{tab:cash}
\end{table*}

We derive a probability of $1.1\times 10^{-3}$ that the two-component model provides a better fit to the data (with respect to the constant polarization model) by random chance. The method for evaluating the statistical significance of the  unbinned log-likelihood analysis is reported in Appendix~\ref{appendix:signifA}.
We have checked the results of the unbinned log-likelihood study by adopting a binned analysis and $\chi^2$ statistics. Details of this check are reported in Appendix \ref{appendix:signifA}. The reduced $\chi^2$ for the two-component model is 1.1 for 13 time bins.

The chosen two-component model assumes that the polarization variability is uncorrelated with the flux variability that we observe.
We also tried a slightly different model, still consisting of a steady plus a variable component. We assumed that the component responsible for the polarization variability also causes flux variability, while the constant polarization component gives a steady contribution to the X-ray flux (flux-correlated two-component model).
In this model we added a parameter: the contribution of the constant component (component 1) to the count rate (r$_1$). By definition, r$_1$ cannot surpass the minimum source counting rate measured during the 2022 October observing period. In this model, there is no longer a degeneracy between $R_1$ and $\Pi_1$, or between $R_2$ and $\Pi_2$. Results of the fit are reported in Table \ref{tab:cash}. Note that the value of the r$_1$ parameter that minimizes the $C$ estimator is at the upper bound of its allowed range (the minimum source counting rate during the observation).

The flux-correlated two-component model and the two-component model are not nested models, so we cannot compare these models directly. From the $\Delta C$ for the flux-correlated
two-component model with respect to the steady polarization model, and from $\Delta C$ for the two-component model with respect to the steady polarization model, we obtain that the probability of choosing the flux-correlated two-component model  by chance from  a
sample of  events with steady polarization is about 1.8 times lower than the probability to choose the two-component model by chance.\\

\subsubsection{X-ray polarimetry -- 2023 August campaign}
\begin{figure*}[tb]
\centering
\begin{tabular}{lll}
\includegraphics[width=0.31\textwidth]{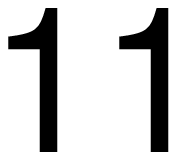} &
\includegraphics[width=0.31\textwidth]{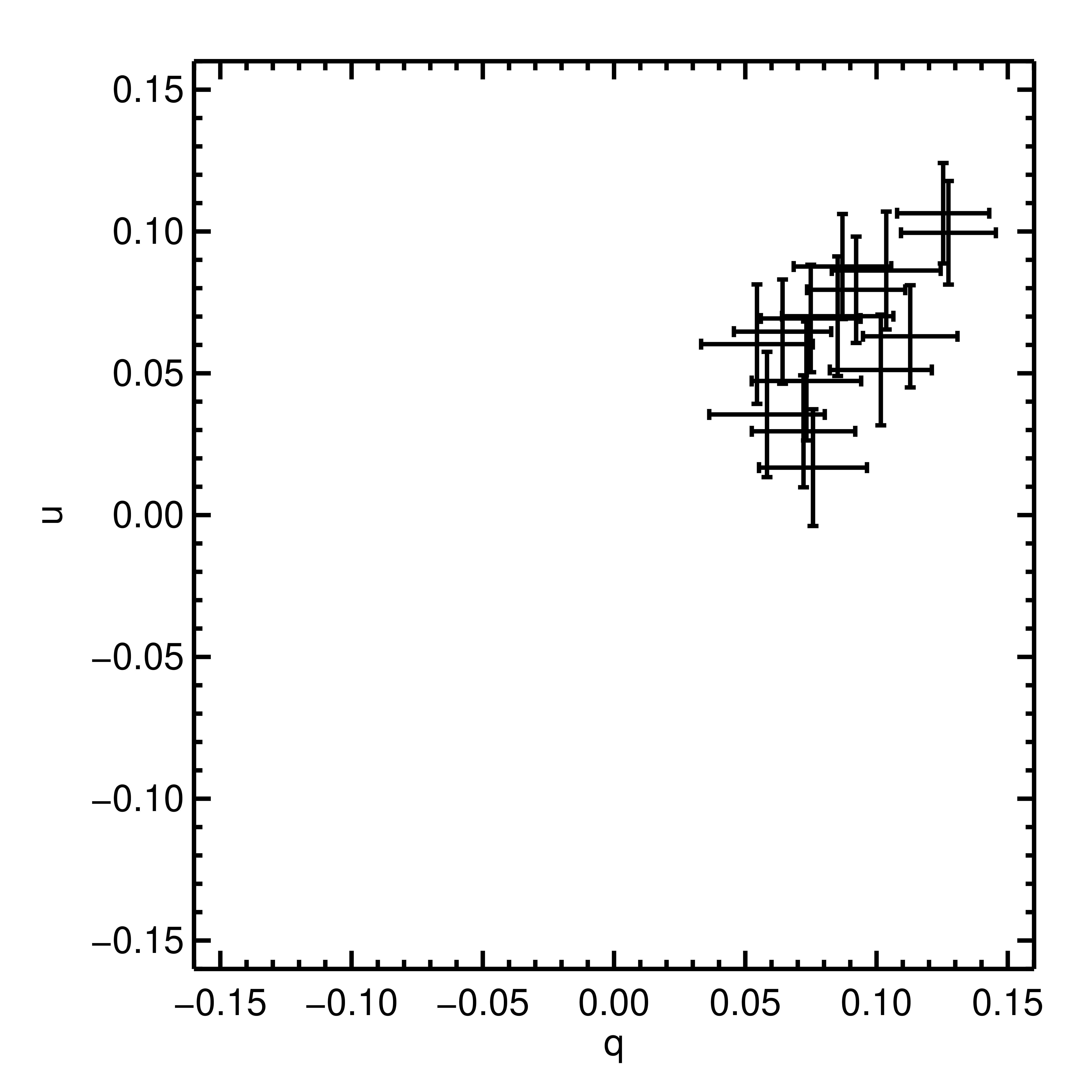} &
\includegraphics[width=0.31\textwidth]
{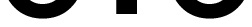} \\
\end{tabular}
\caption{Polarimetric analysis of the 2023 August IXPE data. Fit was performed on the $q$ and $u$ Stokes light curves. Left panel: $q$ and $u$ light curves, fit with a constant polarization model. Central panel: $q$ vs.\ $u$ scatter plot for the 2023 August period.  Right panel: Reduced $\chi^2$ for the constant polarization model as a function of the number of time bins, and associated probability that the fit function is true.}
\label{fig:const_aug2023}
\end{figure*}
\begin{figure}[tb]
\centering
\includegraphics[width=0.5\textwidth]{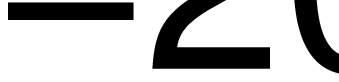}
\caption{Result of search for rotating EVPA for the 2023 August data. We report the minimum of the log-likelihood estimator $C$ for each candidate value of the rotation rate. Model consists of three components (with a rotating plus a counter-rotating component). Results are shown for staggered windows of integration, with size 2.66 d, each shifted from the previous one by 0.88 d.  For increasing window identifier, $C$ is increased by multiples of 30.}
\label{fig:dcash2023}
\end{figure}

During the 2023 August pointing, IXPE detected an average PD of $12.4\;\!\%\pm0.7\;\!\%$ 
and an EVPA of $20\degr\;\!\pm\;\!2\degr$.
The measurements of $q$ and $u$ Stokes parameters as a function of time are reported in Figure \ref{fig:const_aug2023}. The fit with a constant polarization model is reported as well. The $\chi^2$ binned analysis shows that the polarization cannot be considered constant with time.\\
We searched for a rotating component adopting the two component model, with negative results. In fact the two component model produces a circular pattern in $q$, $u$ space, while data show an elongated structure (Figure \ref{fig:const_aug2023}, central panel).
Finally, we tried the three-component model (the new model has a constant component and two counter-rotating components, see Appendix \ref{appendix:threecomp}).
this model produces an ellipse in the $q$,$u$ plane, therefore it could reproduce the elongated pattern observed in August 2023.
The three component model has two more parameters with respect to the two component model: the ratio $\frac{R_2^*}{R_2}\frac{\Pi_2^*}{\Pi_2}$ (where $R_2^*$ is the relative flux of the counter-rotating component, and $\Pi_2^*$ is its PD), and the phase of the counter-rotating component (see Appendix \ref{appendix:threecomp} for detail).
In light of the pronounced flux variability during the observation, we have also searched for rotating components by dividing the sample into sub-intervals.
We integrated data in windows whose lengths are half of the total observing time, and tried four time bins, shifting the bin by one-third of its size, to search for transient phenomena. Results for this staggered analysis are displayed in Figure \ref{fig:dcash2023}. 
We found a signal ($\Delta C=26.6$ for 4 degrees of freedom, plus the rotation-rate parameter, relative to the constant polarization model) at a rotation rate of $5.2\pm0.1$ turn~d$^{-1}$ (one turn is a $360^\circ$ rotation of the EVPA) for an integration window of length $\Delta t=2.66\;\!{\rm d}$,
starting 1.77\;\!d from the beginning of the observation. We report the results of this fit in Table \ref{tab:cash_counter}.
\begin{table*}
    \centering
    \begin{tabular}{cccccc}  \cline{1-2} \cline{5-6} 
    \noalign{\vskip\doublerulesep
         \vskip-\arrayrulewidth} 
     \cline{1-2} \cline{5-6}  
    \multicolumn{2}{c}{steady polarization model} & & & \multicolumn{2}{c}{three-component model with two counter-rotating components}        \\
      C   & $-187.4$ & & & C & $-214.0$ \\
      $\Pi\ (\%)$    &  $9.9\pm 1.2$ & & & $R_1\Pi_1\ (\%)$& $9.9\pm1.2$ \\ 
      $\Psi\ ({\rm deg})$    & $19\pm 2$ & & & $\Psi_1\ ({\rm deg})$ & $19\pm4$ \\\cline{1-2}
      & & & & $R_2\Pi_2+R_2^*\Pi_2^*\ (\%)$ & $5.3\pm 1.6$ \\
      & & & & $\Psi_2(t=0)\ ({\rm deg})$ & $65\pm36$ \\
      & & & & $\omega_2\ ({\rm deg\ d}^{-1})$ & $1864\pm 34$\\
      & & & & $\frac{R_2\Pi_2}{R_2\Pi_2+R_2^*\Pi_2^*}\ (\%)$ & $55\pm 17$ \\
      & & & & $\Psi_2^*(t=0)\ ({\rm deg})$ & $3\pm36$ \\          
     \cline{5-6} \\
    \end{tabular}
    \caption{Parameters and log-likelihood minima for the steady polarization model and for the three-component model
    with two counter-rotating components for the 2023 August observation of 1ES~1959+650 with IXPE.
    $C$ is the log-likelihood minimum, $\Pi$ is the PD, and $\Psi$ is the EVPA of the steady polarized component. The three-component model with two counter-rotating components has 7 parameters: $R_1\Pi_1$ is the product of the relative flux with the PD of component 1, $\Psi_1$ is the EVPA of component 1; $\Psi_2(t=0)$ and $\Psi_2^*(t=0)$ are the EVPAs of rotating and counter-rotating components at the beginning of the observation, respectively, and $\omega_2$ is the angular velocity of the EVPA of the rotating and counter-rotating components.
    The other two parameters of the three-component model are $R_2\Pi_2+R_2^*\Pi_2^*$ and $\frac{R_2\Pi_2}{R_2\Pi_2+R_2^*\Pi_2^*}$, 
    where $R_2$ and $R_2^*$ are the relative fluxes of the rotating and counter-rotating components, respectively, and $\Pi_2$ and $\Pi_2^*$ are the PD of the rotating and counter-rotating components (see Appendix \ref{appendix:threecomp}, eq. \ref{eq:vers_repres_threecomp}).}
    \label{tab:cash_counter}
\end{table*}
We note that, when
integrating data within the same window, there is another signal ($\Delta C$=20.5) for a rotation rate  of $1.9\pm0.1$ turn~d$^{-1}$.

We have validated the unbinned log-likelihood analysis with a binned $\chi^2$ test, as reported in Appendix~\ref{appendix:signifB}.
The chance probabilities are $\le$ 4.0$\times 10^{-3}$ and 6.5\;\!\% for the candidate frequencies at 5.2$\pm$0.1 and 1.9$\pm$0.1 turn~d$^{-1}$, respectively. We report the evaluation of the significance of the unbinned log-likelihood fit of the three-component model in Appendix~\ref{appendix:signifB}.
\begin{figure*}[tb]
\centering
\begin{tabular}{lll}
\includegraphics[width=0.31\textwidth]{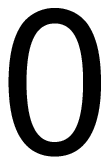} & 
\includegraphics[width=0.31\textwidth]{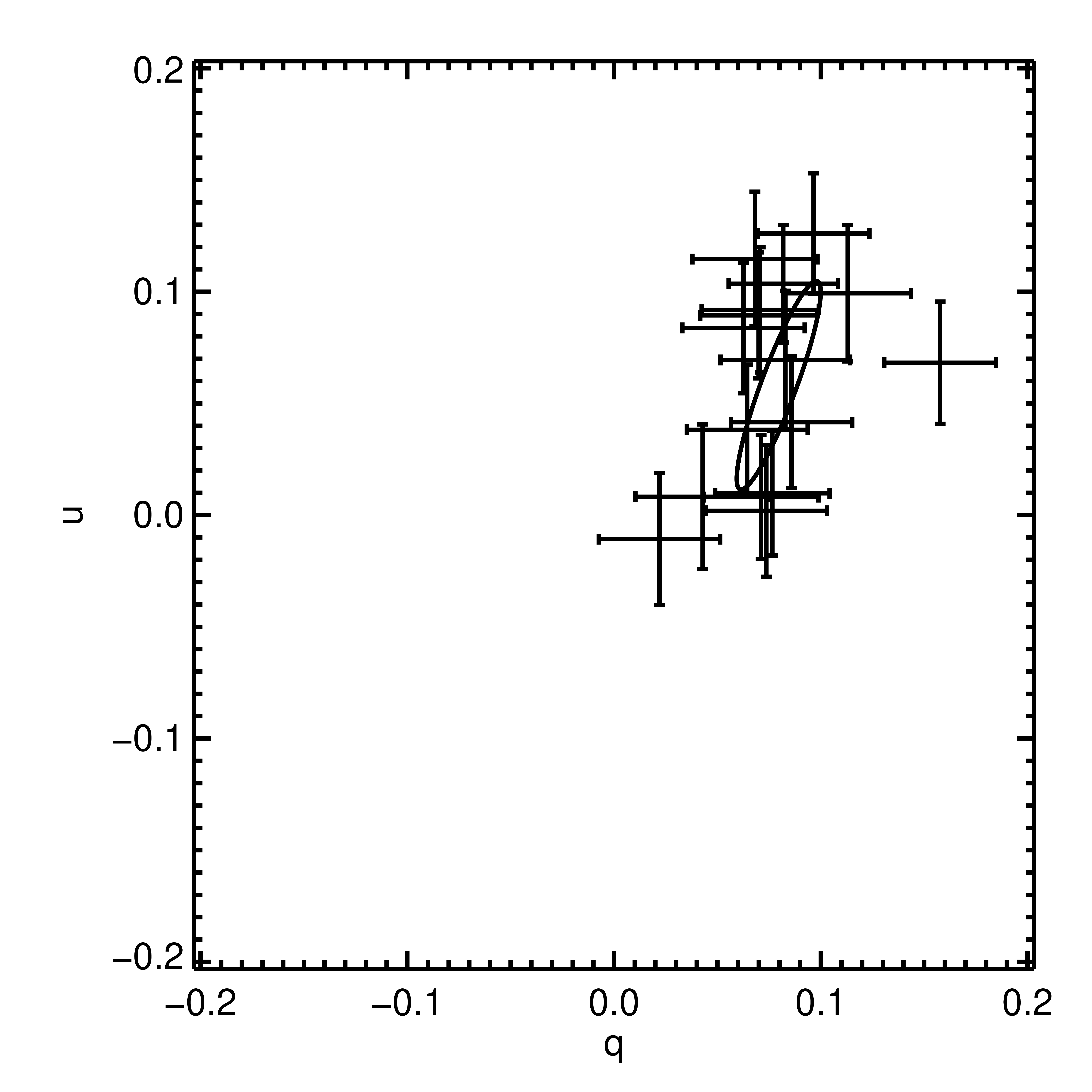} &
\includegraphics[width=0.31\textwidth]{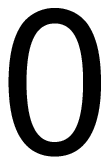} \\
\end{tabular}
\caption{Left panel: Folded polarimetric light curve of X-ray $q$ and $u$ Stokes parameters, for a folding frequency of 5.2 $\pm$ 0.1 d$^{-1}$. Signal is accumulated for the entire window 3 of the 2023 August IXPE pointing (see text); the continuous line is the three-component model fitting function, while the dotted curve is the constant polarization model. Central panel: $q$ vs $u$ plot for the folded signal; the ellipse (continuous line) is the three-component model fitting function. Right panel:  Folded polarimetric light curve of X-ray $q_{\rm rotated}$ and $u_{\rm rotated}$ Stokes parameters, rotated clock-wise by 70$\degr$ (the rotation angle corresponds to the angle of the ellipse (of the three-component model) major axis with respect to the $q$ reference axis.)}
\label{fig:phased_pol_lc}
\end{figure*}

We report in Figure \ref{fig:phased_pol_lc} (left panel)
the X-ray PD and EVPA signals folded for a frequency of 5.2 $\pm$ 0.1 turn~d$^{-1}$, accumulated for the entire window 3; and in Figure
\ref{fig:phased_pol_lc} (central panel) the corresponding $q$ vs.\ $u$ plot.\\

The right panel of Figure \ref{fig:phased_pol_lc}
displays the folded polarimetric light curves in a 70$\degr$ rotated $q,u$ space. It shows that the modulated signal is
almost uni-dimensional. We tested the hypothesis of an uni-dimensional oscillation by fixing the $\frac{R_2\Pi_2}{R_2\Pi_2+R_2^*\Pi_2^*}$
parameter to 0.5. With this choice,  the three-component model describes a line in $q,u$ space, with Stokes parameters oscillating around the center of the line. With this choice, $\Delta\ C\ =-25.3$ for the principal minimum at $5.2\ {\rm d}^{-1}$ rotating frequency; while  $\Delta C\ =-20.5$ for the secondary minimum at $1.9\ {\rm d}^{-1}$. In this case, $\Delta C$ has a $\chi^2$ distribution with 3 degrees of freedom when computed as a function of the EVPA rotation rate. The probability to detect by chance at least two such signals with $\Delta C\ \le \ -20.5$
in the August 2023 observation~is~$\le\ 2.5\times 10^{-4}$.\\

We have also investigated whether the rotation rate could vary with some power of X-ray energy: $\omega=\omega_0\left(\frac{E}{2\ {\rm keV}}\right)^{k}$. For the 2022 October  observation we obtained $k=0.1\pm0.1$; while for the 2023 August  ToO we obtained  $k=0.002\pm0.005$  (90\% c.l.). For both observations, the result is compatible with a rotation rate that does not vary with energy (90\% c.l.).  
We have tested with simulations whether gaps in the observation or the satellite pointing dithering could produce a spurious rotating signal at the frequency range of our analysis, but were unable to reproduce such a signal. Details are given in Appendix~\ref{appendix:signifB}.
\subsubsection{Radio and optical polarimetry}

During the 2022 October campaign, we measured a low degree of polarization of 1ES~1959+650 at 225 GHz: 
$\Pi_R=1.3\pm0.4\;\!\%$ at a position angle of $\psi_R=140\pm9^\circ$.  At optical wavelengths, the flux varied on short timescales, but with a low amplitude of $<$0.1~mag. The average intrinsic R-band polarization of the source was $\Pi_O=4.54\;\!\%\pm0.7\;\!\%$ at EVPA $\psi_O=152\degr\pm6\degr$. All  B, V, R, and I EVPAs are consistent within uncertainties and appear to vary in tandem. The jet direction prior to 2018 was $\mathbf{128\degr\pm13\degr}$ \citep{Weaver2022},
although during the IXPE observations it
was (including uncertainties) $\sim148$-$168\degr$ (see below). Hence, the radio-optical polarization was roughly aligned with the jet axis within the rather large uncertainties. 

While the source was in outburst during the 2023 August campaign, we did not observe significant differences in its radio-optical polarization properties. The flux exhibited similar fast variations at low amplitudes. The radio (10--225\;\!GHz) polarization was in the range $\Pi_R=1\;\!\%$--$4\;\!\%$ with EVPA between $\psi_R=138\degr$ and $159\degr$, with a typical uncertainty $\sigma_{\psi_R}=6\degr$.
The R-band polarization increased from $4\;\! \%$ to $6\;\! \%$ during the IXPE observation, and then returned to $\sim 4 \;\!\%$ afterward. The EVPA varied from 144$^\circ$ to 171$^\circ$.

\begin{figure}[bt]
\centering
\includegraphics[width=0.5\textwidth]{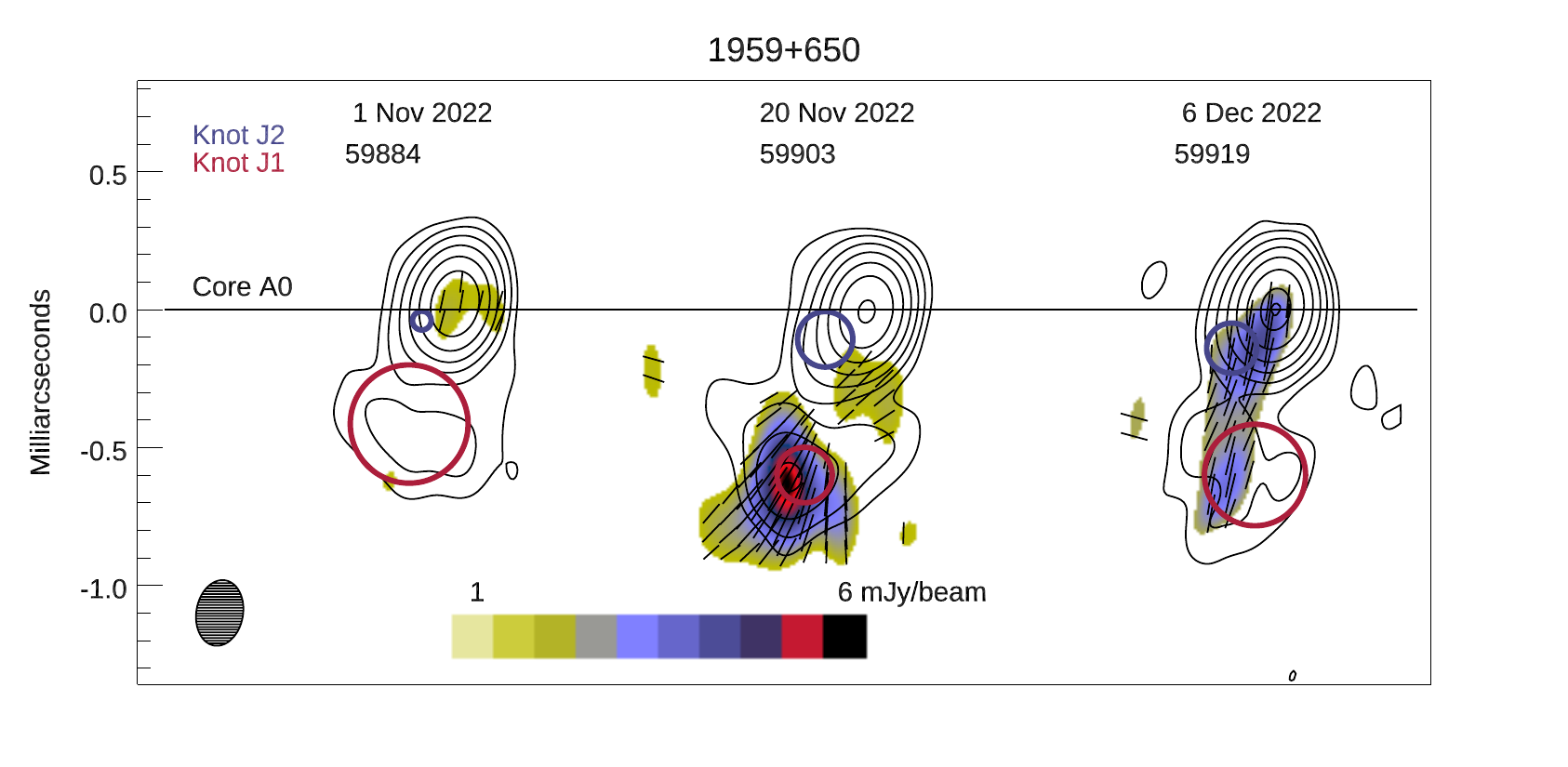}   
\includegraphics[width=0.5\textwidth]{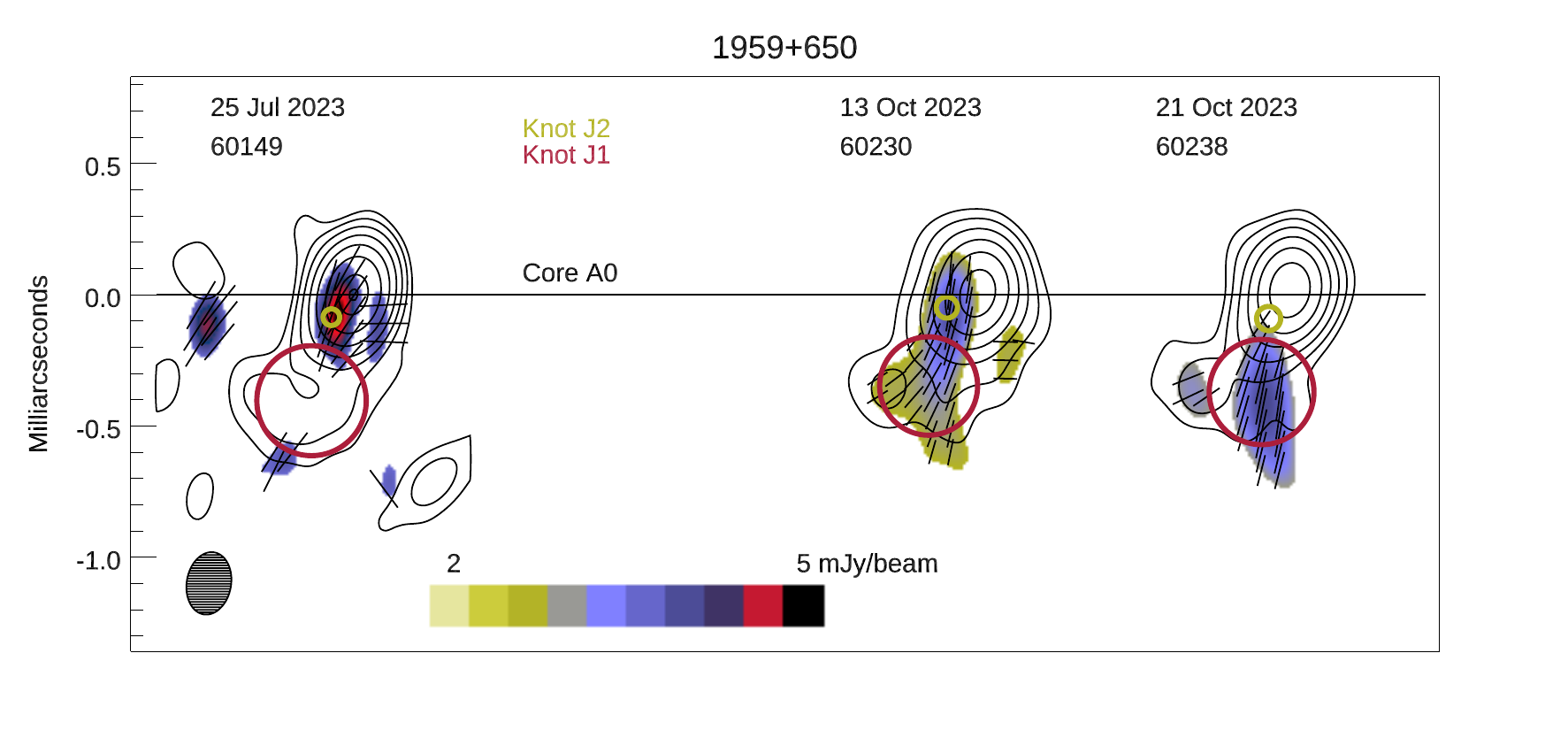}   
\caption{VLBA images at 43 GHz of 1ES~1959+650 at three epochs near each IXPE observation, which occurred in 2022 October 28-31 and 2023 August 14-19. Both calendar dates and MJDs of the images are given. Three emission features present in all images, A0 (the ``core,'') J1, and J2, are marked, with their parameters listed in Table \ref{tab:VLBAknots}. Contours  indicate total intensity in factors of 2, starting at 1.2\% of the peak of 156 mJy/beam in 2022 and 255 mJy/beam in 2023. Color scale corresponds to linearly polarized intensity, with values indicated in each panel, while the polarization angle is denoted by the line segments in regions where polarization is detected. The elliptical restoring beam, with FWHM dimensions of $0.24\times0.17$ mas, with major axis along position angle $-10\degr$, is displayed in the lower left corner of each panel.}
\label{fig:VLBAimages}
\end{figure}

\subsubsection{ Millimeter-wave Imaging}

The VLBA observes 1ES~1959+650 roughly monthly at a 
frequency of 43 GHz as part of the Blazars Entering 
the Astrophysical Multi-Messenger Era (BEAM-ME) 
monitoring program \footnote{www.bu.edu/blazars/BEAM-ME.html}. In Figure \ref{fig:VLBAimages} we present two
sets of three images each, constructed from data obtained
near each of the two IXPE epochs. The angular resolution is of order 0.15 milliarcseconds (mas), which translates to 0.14 pc projected on the sky for a Hubble constant of 70 km s$^{-1}$ Mpc$^{-1}$. The imaging 
process involves a number of calibration steps, 
followed by iterations of image construction and self-calibration; see \citet{Jorstad2017} for a thorough description of the analysis procedures. 
The object is weak at 43 GHz, limiting the dynamic range of the images such that the lowest contours are affected by noise. In order to represent the main features of each image, we fit the $u$-$v$ data
with three components of circular Gaussian brightness distributions. Table \ref{tab:VLBAknots} lists the parameters of these components at each epoch. The brightest feature, A0, is referred to as the ``core.'' In blazars, the core is considered stationary and near the upstream end of the jet. Knots (J1 and J2 in 1ES~1959+650) downstream of the core can either be quasi-stationary (common in BL Lac objects) or move away from the core \citep{Jorstad2017}. The only motion apparent in the images of Figure \ref{fig:VLBAimages} is a downstream shift in position of knot J1 between 2022 November 1 and 20 that does not continue to the next epoch 16 days later.

The jet direction (defined by a line between the centers of the core and knot J1) is $163\degr\pm5\degr$ during the 2022 epochs and $155\degr\pm2\degr$ in 2023. This is similar to the optical polarization angle. The degree of polarization of knot J1 tends to be high --- up to $36\pm16\%$ --- although the uncertainties are large. The 43 GHz polarization angles of the core and knots, when detected, are within $\pm10\degr$ of the jet direction as well.

\begin{deluxetable*}{llrrrrrrr}
\label{tab:VLBAknots}
\singlespace
\tablecolumns{9}
\tablecaption{Parameters of Knots in 43 GHz VLBA Images \label{JParm}}
\tabletypesize{\footnotesize}
\tablehead{
\colhead{Date}&\colhead{MJD}&\colhead{ID}&\colhead{Flux Density}&\colhead{Distance from A0}&\colhead{Position Angle}&\colhead{Diameter}&\colhead{$\Pi$}&\colhead{$\chi$}\\
\colhead{}&\colhead{}&\colhead{}&\colhead{mJy}&\colhead{mas}&\colhead{degrees}&\colhead{mas}&\colhead{\%}&\colhead{degrees}\\
\colhead{(1)}&\colhead{(2)}&\colhead{(3)}&\colhead{(4)}&\colhead{(5)}&\colhead{(6)}&\colhead{(7)}&
\colhead{(8)}&\colhead{(9)}
}
\startdata
2022/11/01& 59884&A0&117$\pm$15&   0.000&\nodata&0.043$\pm$0.005&1.5$\pm$1.1&169$\pm$11\\
& &              J2&6$\pm$3&0.14$\pm$0.01& 99$\pm$9&0.068$\pm$0.007&$<$1.6&\nodata \\
& &              J1&37$\pm$9&0.42$\pm$0.04&168$\pm$6&0.43$\pm$0.04&$<$16&\nodata \\
2022/11/20& 59903&A0&125$\pm$17&   0.000&\nodata&0.040$\pm$0.005&$<$0.6&\nodata \\
& &               J2&35$\pm$7  &0.15$\pm$0.01&136$\pm$8&0.20$\pm$0.01&$<$1.9&\nodata \\
& & J1&33$\pm$9 &0.68$\pm$0.05&156$\pm$6&0.20$\pm$0.02&36$\pm$16&151$\pm$10 \\
2022/12/06& 59919&A0&149$\pm$10&   0.000&\nodata&0.030$\pm$0.004&2.2$\pm$1.5&173$\pm$5 \\
& &               J2&24$\pm$6&0.15$\pm$0.01&126$\pm$8&0.18$\pm$0.01&7.1$\pm$4.8&169$\pm$11 \\
& &               J1&26$\pm$9&0.60$\pm$0.04&165$\pm$6&0.37$\pm$0.02&43$\pm$24&167$\pm$16 \\
2023/07/25& 60149&A0&232$\pm$12&   0.000&\nodata&0.022$\pm$0.004&1.5$\pm$0.8&138$\pm$10 \\
& &               J2&38$\pm$9&0.110$\pm$0.013&141$\pm$10&0.063$\pm$0.006&3.6$\pm$1.5&155$\pm$8 \\
& &               J1&73$\pm$11&0.40$\pm$0.04&151$\pm$9&0.42$\pm$0.06&$<$21&\nodata \\
2023/10/13& 60230&A0&111$\pm$10&   0.000&\nodata&0.039$\pm$0.005&$<$1.1&\nodata \\
& &               J2&70$\pm$8&0.14$\pm$0.01&152$\pm11$&0.081$\pm$0.007&7.1$\pm$4.6&169$\pm$10 \\
& &               J1&42$\pm$12&0.40$\pm$0.01&149$\pm$9&0.38$\pm$0.04&23$\pm$17&145$\pm$21 \\ 
2023/10/21&  60238&A0&122$\pm$15&  0.000&\nodata&0.025$\pm$0.006&$<$0.8&\nodata \\
& &                J2&62$\pm$9&0.15$\pm$0.02&160$\pm$11&0.09$\pm$0.01&$<$6.5&\nodata \\
& &                J1&43$\pm$11&0.44$\pm$0.04&148$\pm$8&0.40$\pm$0.05&29$\pm$21&155$\pm$21 \\
\enddata
\end{deluxetable*}

\subsection{Energy Spectra}
\subsubsection{2022 October campaign}
\begin{figure*}[tb]
\centering
\begin{tabular}{l}
\includegraphics[width=0.7\textwidth]{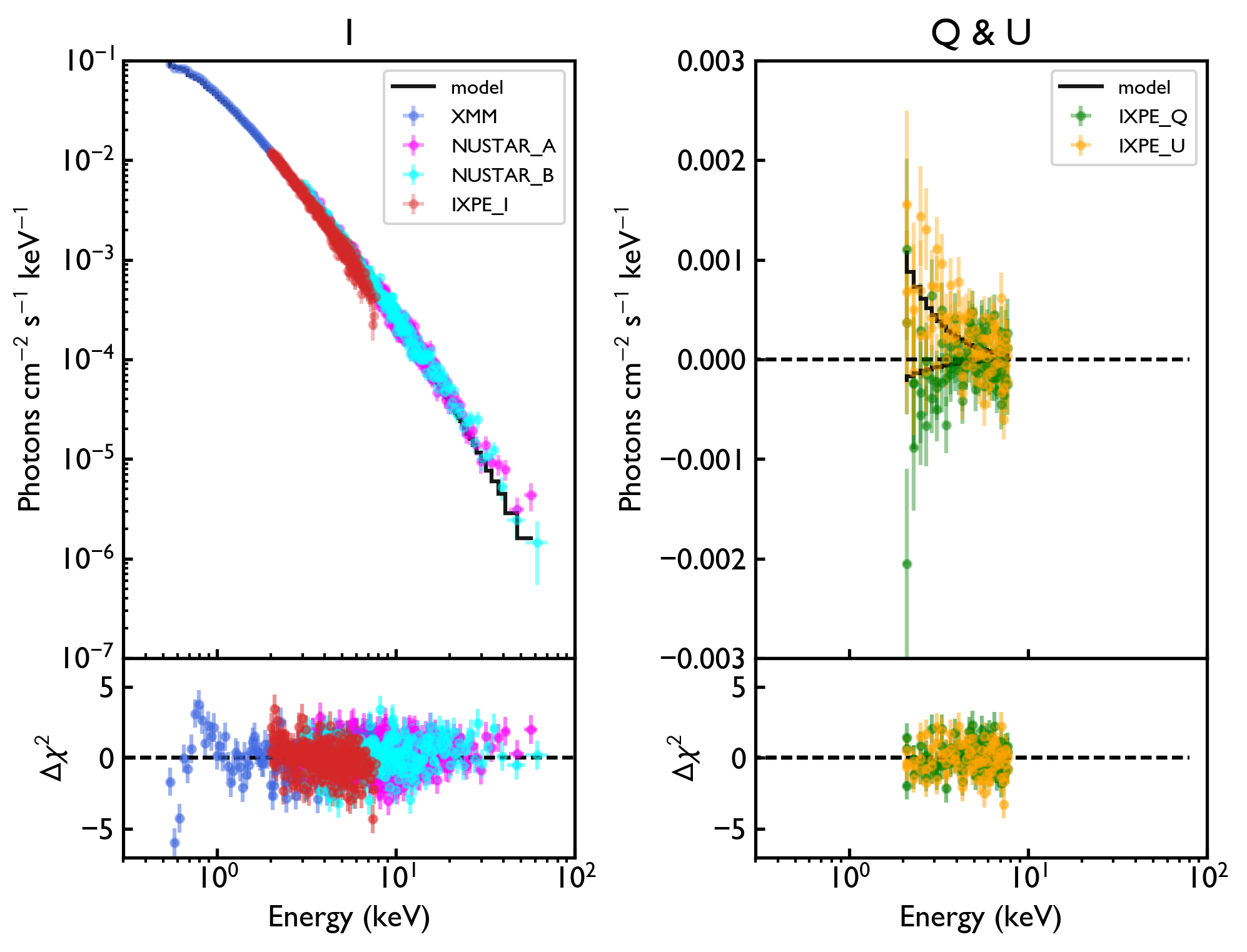}  \\
\includegraphics[width=0.7\textwidth]{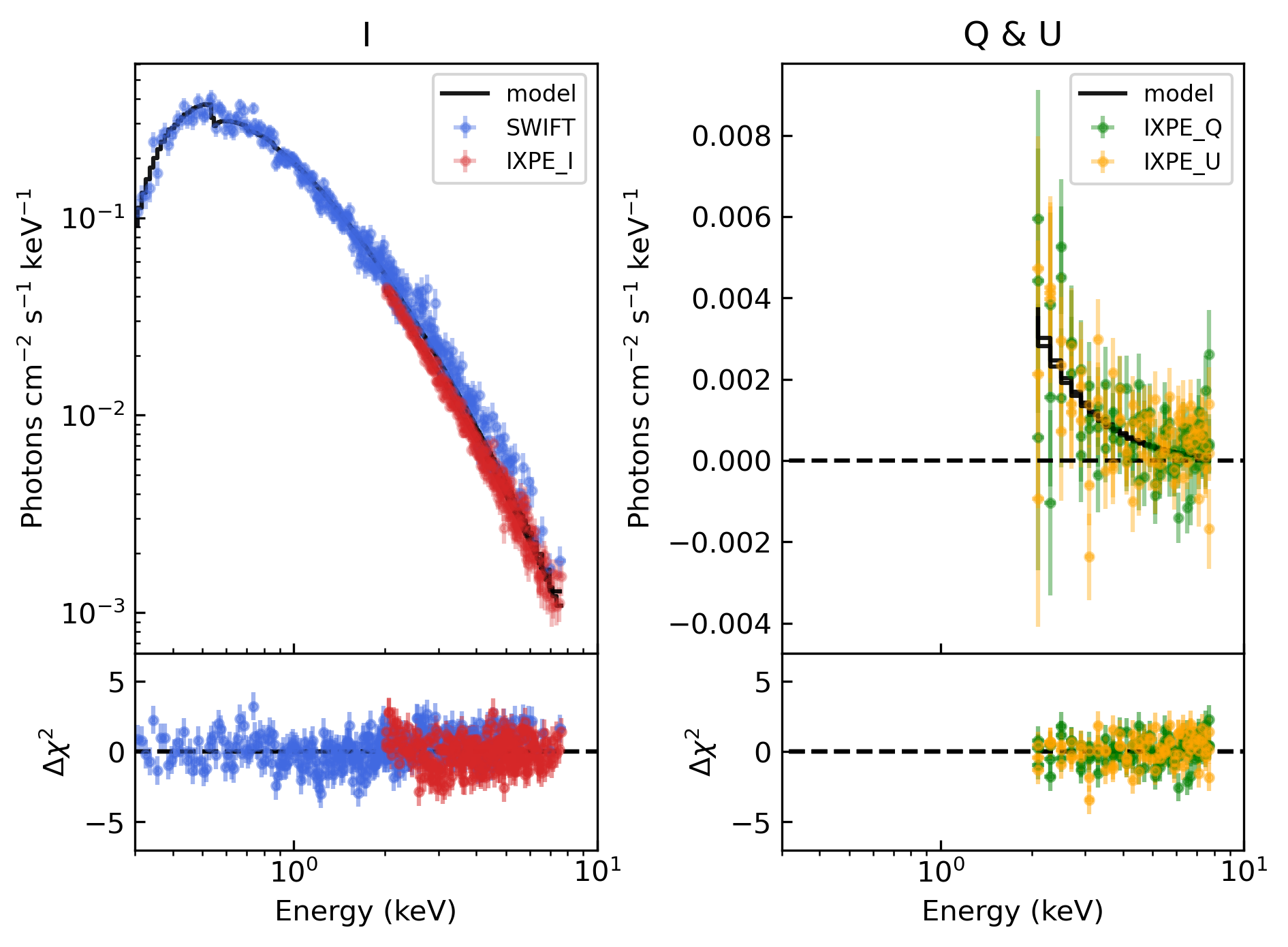} 
\end{tabular}
\caption{
\textit{Top:} Combined X-ray spectrum of 1ES~1959+560 during the 2022 October multi-wavelength campaign, including Swift-XRT, IXPE, XMM-Newton, and NuSTAR data. \textit{Bottom:} Combined X-ray spectrum during the 2023 August campaign, including Swift-XRT and IXPE data integrated for 1.0 d around the peak emission on MJD 60172.}
\label{fig:broadspectra2022_2023}
\end{figure*}

The X-ray spectrum obtained during the 2022 October campaign is shown in Figure \ref{fig:broadspectra2022_2023}. Data from Swift-XRT, IXPE, XMM-Newton, and NuSTAR are reported. An absorbed log-parabola model was fit to the data. Fit parameters are reported in Table \ref{table:spefit_oct2022}. We obtained an un-absorbed flux of $(8.61\pm0.04)\times10^{-11}$ erg cm$^{-2}$ s$^{-1}$ in the 2$-$8 keV band (90\% c.l.).\\

\subsubsection{2023 August campaign}
Swift observed the blazar 6 times during the 2023 August IXPE observation. We derived the spectrum by integrating IXPE data over 1.0\,d around the peak emission on MJD 60172. The spectrum thus obtained is displayed in Figure \ref{fig:broadspectra2022_2023}, while the fit parameters are reported in Table \ref{table:spefit_aug2023}. We obtained an un-absorbed flux of $(31.4\pm0.6)\times10^{-11}$ erg cm$^{-2}$ s$^{-1}$  (90\% c.l.) in the 2--8 keV band.\\
\begin{table*}
    \centering
    \begin{tabular}{c c c c c c c c c} \\ \hline \hline
$\frac{\chi^2}{\rm dof}$  & dof &N$_{\rm H}\ (10^{21}\ {\rm cm}^{-2})$ &$\Pi_{\rm X}\ (\%)$  & $\Psi_{\rm X}\ ({\rm deg})$ & $\alpha$   & $\beta$ &$E_{\rm pivot}$\ (keV)& $N\ (10^{-3}\ {\rm keV}^{-1}\ {\rm cm}^{-2})$   \\
1.22                   & 1214 &1.01&9.8$\pm$1.2& 50$\pm$4 &2.504$\pm$0.005&0.323$\pm$0.005&  5  &1.403$\pm$0.005 \\ \hline
    \end{tabular}
 \caption{Log-parabola model X-ray spectral fit of the combined 2022 October IXPE, NuSTAR, XMM-Newton, and Swift-XRT data of 1ES~1959+650. The number of degrees of freedom are denoted with dof; the column density N$_H$ and pivot energy are held constant. 
 Not shown in the table are the inter-calibration factors with respect to XMM-Newton; for NUSTAR PFMA and PFMB: $1.14\pm0.01$ and $1.14\pm0.01$, respectively; for IXPE DU1, DU2, DU3: $1.033\pm0.005$, $0.991\pm0.005$, and $0.936\pm0.005$, respectively.}
 \label{table:spefit_oct2022}
 \end{table*}
\begin{table*}
    \centering
    \begin{tabular}{c c c c c c c c c} \\ \hline \hline
$\frac{\chi^2}{\rm dof}$  & dof &N$_{\rm H}\ (10^{21}\ {\rm cm}^{-2})$ &$\Pi_{\rm X}\ (\%)$  & $\Psi_{\rm X}\ ({\rm deg})$ & $\alpha$   & $\beta$ &$E_{\rm pivot}$\ (keV)& $N\ (10^{-3}\ {\rm keV}^{-1}\ {\rm cm}^{-2})$   \\      
1.12                   & 787 &1.01&12.6$\pm$1.3& 22$\pm$3 &2.98$\pm$0.02&0.80$\pm$0.02&  5   &  4.65$\pm$0.06 \\ \hline
 \end{tabular}
 \caption{X-ray spectral fit of a log-parabola model to the combined 2023 August IXPE and Swift-XRT data for 1ES~1959+650. The number of degrees of freedom is denoted with dof; the column density N$_H$ and pivot energy  are held constant. The inter-calibration factors, not shown in the table are with Swift-XRT as reference: IXPE DU1, DU2, DU3: $0.873\pm0.009$, $0.829\pm0.009$, $0.793\pm0.009$, respectively.}
 \label{table:spefit_aug2023}
 \end{table*}

\section{Discussion} \label{sec:discussion}
Our X-ray observations found 1ES~1959+650 to be at an intermediate flux level in 2022 October and in outburst in 2023 August. During the second campaign, the X-ray spectrum
gradually softened as the flux declined, except for a brief period of hardening corresponding to the flux peak on MJD 60172. 
This can be interpreted as gradual cooling of the electron population, interrupted by an episode
of enhanced acceleration of electrons.

Optical photometry in 2023 August measured the brightness to be 1.2$-$1.5 mag higher than in 2022 October.
In contrast, the millimeter-wave flux did not vary significantly between the two epochs.
The Swift-UVOT flux in 2023 August
did not exhibit the flaring episode that peaked on MJD 60172, as observed in the X-ray range with both IXPE and Swift-XRT.
Swift-UVOT observations with optical filters only showed a weak flux peak on MJD 60172.

For both periods, the optical PD was in the $4\;\!\%-6\;\!\%$ range, and the EVPA fluctuated within 15$\degr$ of
153$\degr$.
 IXPE observations in 2022 October and 2023 August measured the PD and EVPA to change from $9.4\;\!\%\pm1.6\;\!\%$ to $12.4\;\!\%\pm0.7\;\!\%$, 
 and from $53\degr\pm 5\degr$ to  $12.4\degr \pm 0.7\degr$, respectively.
 
Interestingly, the R-band PD in 2023 August increased from $\sim 4\%$ to $\sim6\%$ from the beginning to the end of the IXPE pointing, while the EVPA varied more erratically by $\pm10\degr$. In contrast,  IXPE found higher
X-ray polarization, without any increasing PD trend; see Figure~\ref{fig:const_aug2023}.

Our IXPE measurements have found significant X-ray polarization, with evidence for
a component with rotating EVPA during both observations.
We have found that the EVPA rotation rate does not vary with X-ray energy.
Electrons incoherently gyrating in a magnetic field follow the field lines, with EVPA orthogonal to the local magnetic field direction as projected on the sky. Plasma moving in a helical magnetic field will produce synchrotron radiation with a rotating EVPA when observed close to the 
symmetry axis of the helix.

We have found some evidence for a rapidly rotating ($\sim$5.2 turn~d$^{-1}$, 0.22\;$\%$ chance probability) component starting 1.77\;\!d from the beginning of the 2023 August observation, which 
suggests that the rotation began with the peak in activity of the source. We found this signal only by 
adding a counter-rotating component to the two-component model.
In general, a rotating plus a counter-rotating component describe an ellipse in the $(q,u)$ plane (see Appendix \ref{appendix:threecomp}).
In our case, the two components had almost the same flux, so that the model location in the $(q,u)$ plane lies
along a nearly straight line.
Our three-component model reproduces the X-ray polarimetric variability observed for window 3 of the 2023 August IXPE pointing. However, we cannot establish if there was a counter-rotating component responsible for the observed emission, or if, on the contrary,
the counter-rotating component only allowed us to describe a one-dimensional oscillation in the $(q,u)$ plane caused by a different emission mechanism.
A further rotating signal ($\sim$ 1.9 turn~d$^{-1}$) with lower significance is found for the same exposure time during the 2023 August campaign.
\begin{figure}[bt]
\centering
\includegraphics[width=0.35\textwidth]{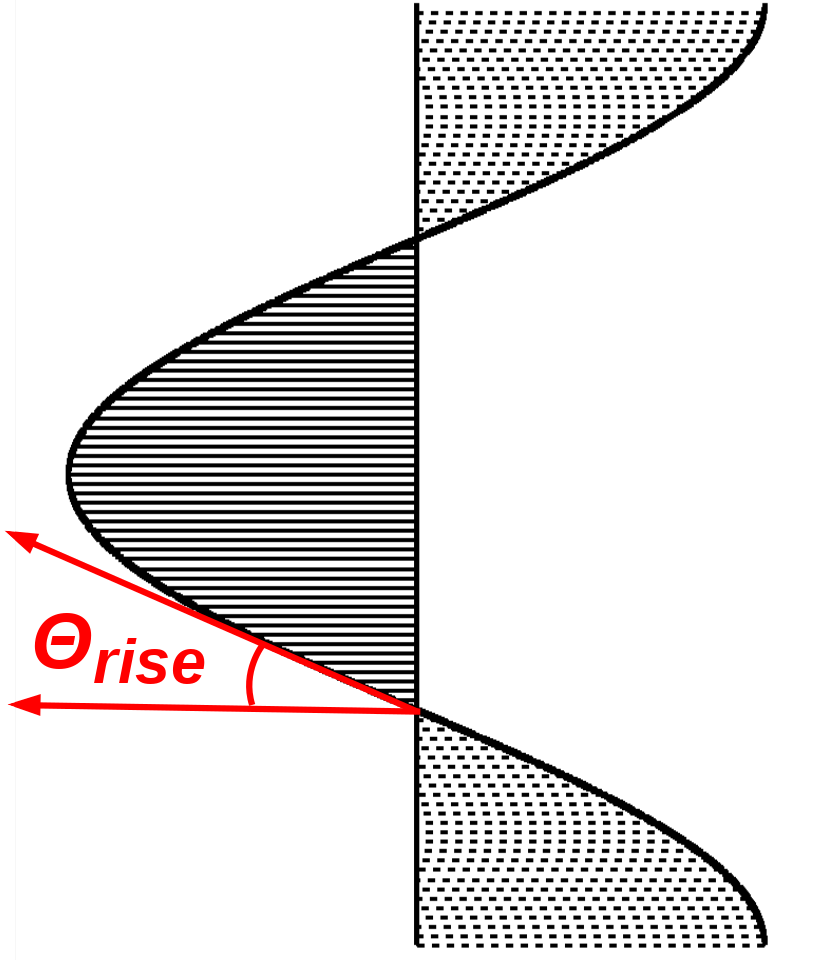}   
\caption{Helix viewed at $90^\circ$ to the axis, illustrating the definition of the rising angle $\Theta_{\rm rise}$, the regions seen from above (shading by continuous lines), and regions seen from below (dashed lines). The jet axis corresponds to the vertical line.}
\label{fig:helicoid}
\end{figure}
%
%
Two previous papers \citep{DiGesu2023,Kim2023} have discussed how helical motion of plasma could explain the phenomenology observed in Mrk~421.
For a helix, we can define the instantaneous rising angle ($\Theta_{\rm rise}$ of the plasma moving along the helical path (see Figure  \ref{fig:helicoid}),  while the angle  formed by the instantaneous velocity vector of plasma moving along a helical path and the symmetry axis of the helix
is denoted by $\Theta_{\rm rise} ^{\rm compl}$. (By definition, $\Theta_{\rm rise}^{\rm compl}=\frac{\pi}{2}-\Theta_{\rm rise}$.)
We note that for the rotating signal to be detected along the entire helical orbit, $\Theta_{\rm rise}^{\rm compl}$ should be smaller than the critical jet viewing angle ${1}/{\Gamma_{\rm bulk}}$. Moreover, the observer's viewing angle ($\Theta_{\rm view}$, measured with respect to the axis of the helix) should be lower than  $\Theta_{\rm rise}^{\rm compl}$ for the Stokes parameters to execute a circular shape in the $q,u$ plane:
\begin{equation}
 \biggr\{ \begin{array}{c}
 \Theta_{\rm rise} ^{\rm compl} \ < {1}/{\Gamma_{\rm bulk}}\\
 \Theta_{\rm view} \ <  \Theta_{\rm rise} ^{\rm compl}
\end{array} \ .
 \label{eq:thetarise_minor}
\end{equation}
This implies that the plasma's helical path is elongated along the jet axis.\\

The EVPA rotation observed in the 2022 October X-ray data of 1ES~1959+650 can be explained by the hypothesis corresponding to Equation~(\ref{eq:thetarise_minor}). We note, however that the observed X-ray polarization  evolution can be interpreted also with the flux-correlated two-component model. With this modeling, we obtained that the fitted flux of the constant component corresponds to
the maximum allowed value (see Table~\ref{tab:cash}), possibly indicating that the line of sight was not inside the source emitting cone for the entire observation: $\Theta_{\rm view} \ \sim  \Theta_{\rm rise} ^{\rm compl}$, or, alternatively, that the typical opening angle of the helix is comparable to (or slightly larger than) the emitting cone aperture: $\Theta_{\rm rise} ^{\rm compl} \ \sim {1}/{\Gamma_{\rm bulk}}$.

The peculiar rotation of the EVPA observed in 2023 August cannot be explained with Equation~(\ref{eq:thetarise_minor}). We also could not
reproduce the observed behavior with  the hypothesis that the our line of sight is outside $\Theta_{\rm rise} ^{\rm compl}$: 
\begin{equation}
 \biggr\{ \begin{array}{c}
 \Theta_{\rm rise} ^{\rm compl} \ < {1}/{\Gamma_{\rm bulk}}\\
 \Theta_{\rm view} \ >  \Theta_{\rm rise} ^{\rm compl}
\end{array} \ .
 \label{eq:thetarise_minor_without_tview}
\end{equation}
The evolution of rotated Stokes parameters in Figure~\ref{fig:phased_pol_lc} suggests that the polarimetric oscillation takes place in a single dimension.
Alternatively, we have the unlikely case of two almost exactly counter-rotating (and of similar flux) components combining to give rise to the observed polarimetric X-ray signal.

EVPA rotation components could be a sign of a stochastic process controlling the direction of the magnetic field. Rotations that vary the EVPA about a ``preferred'' value can be explained also by turbulent plasma whose field is partially ordered by a shock or by a helical component \citep{Marscher2014}.
\section{Interpretation} \label{sec:interpretation}
The two campaigns centered on IXPE observations of 1ES~1959+650 found different X-ray PDs and EVPAs, which indicates that the X-ray polarization is not simply associated with a fixed jet direction.
We propose that the polarized X-ray emission was associated with localized regions during each of the two campaigns, although there were substantial differences. 
During both observations, we detected evidence for components with rotating EVPAs. Over the intermediate activity period observed during the 2022 October campaign, the EVPA rotation velocity was low ($\sim$ 0.3 turn~d$^{-1}$). Our modeling of the IXPE data from the 2023 August campaign, during a major outburst of the source, instead found a fast rotation velocity ($\sim$5.2 turn~d$^{-1}$, detected when integrating data over 2.66d).
Another difference between the two campaigns is that, in 2022 October, we only needed a constant and a single rotating component, while in the 2023 August campaign, a counter-rotating component with almost the same intensity as the rotating one is required for the rotating EVPA signal to be detected, resulting in an elongated elliptical path in $q$, $u$ space.
There may be an analogy with solar coronal loop observations \citep[see, e.g., ][ and references therein]{nakariakov2016}:  both rapidly decaying and undamped oscillations are detected. For coronal loops, continuous excitation of the oscillating system is invoked to explain the latter case.

Regarding optical polarization, our data confirm the presence of a dominant and persistent polarized emission component with a fixed EVPA direction,
with other weak components contributing to minor changes of the EVPA. We observed for the X-ray outburst in 2023 August a gradual rise of the optical PD, and a coincident change of the optical EVPA. Moreover, optical and
UV photometric measurements show that, over long time spans, the optical and X-ray emission both rose in 2023 August with respect to
emission in 2022 October, and that the X-ray flux peak on MJD 60172 was observed in V band as well. Therefore,
a dominant and persistent optical emission component is observed, but
the data suggest that the plasma responsible for the X-ray emission provides a minor contribution to the optical flux.

In contrast with the X-ray emission, the optical polarization vector in 2023 August was, on average,
oriented in a direction similar to that measured in 2022 October, even though the optical flux on 2023 August was double the level of 2022 October. 
This discrepancy can be reconciled by assuming that the X-rays are emitted within small regions where the mean direction of the magnetic field is not perpendicular to the jet axis as measured at 43 GHz,
while optical photons are emitted from larger regions where the average magnetic field is transverse to the jet axis; see below for proposed models.

Regarding the radio flux and polarization, our dataset is rather sparse for the 2022 October period, while for the 2023 August campaign the cadence is daily. The average radio PD is lower than the optical value, while the EVPA almost matches
that at optical wavelengths. This
suggests that the main source of radio flux has the same general origin as the dominant and persistent emission observed at optical wavelengths.

\citealt{2010ApJ...723.1150P} 
measured the apparent speed of moving components near the core at 43~GHz in TeV-detected blazars, finding low apparent speeds (below 1$c$) for 1ES~1959+650, Mrk~421, and Mrk~501, implying a viewing angle $\Theta_{view}\ <\ \frac{1}{\Gamma_{bulk}^2\beta}$, and $\Gamma_{bulk} \sim3\ -\ 5$. We argue for a similar scenario for the X-ray emission (see Equation \ref{eq:thetarise_minor}). At such a narrow viewing angle, the resolution of the VLBA images, $\sim0.2$ milliarcseconds, corresponds to parsec scales when deprojected.

\cite{Weaver2022} display an image of the blazar obtained in 2018 July at 43 GHz, showing the core A0, and knots A1, A2, A3, A4. A line of sight very close to the jet direction explains the rather large apparent opening angle, as well as the bend to the north on kiloparsec scales \citep{Rector2003}, which can be a small bend amplified by projection effects if the viewing angle is near $0\degr$. (The jet direction is obtained by from the position angle of lines connecting the core to the other features; position angles vary from $124\degr\pm8\degr$ to $163\pm17\degr$; \citealt{Weaver2022}.) Our similar high-resolution 43 GHz VLBA images in 2022 and 2023 (Fig.\ \ref{fig:VLBAimages}) contain two features, J1 and J2, southeast of the core, with the inner jet EVPA oriented in the jet direction, indicative of a mean magnetic field that is transverse to the jet on parsec scales.

Polarization measurements at 10.45, 17, 43, and 225 GHz are core-dominated, and they represent the average jet polarization on parsec scales (the innermost jet region is obscured by self-absorption at the lower frequencies), with EVPA aligned with the jet direction. We surmise that the observed polarization degree is low because of a dominant turbulent component of the magnetic field, with the electric vector polarization angle reduced by vector-averaging over the broad jet apparent opening angle. The high-resolution images of the jet 
(see Fig. \ref{fig:VLBAimages}) confirm that the EVPA is parallel to the jet. Feature J1 usually has a high polarization degree, indicating a moderate degree of ordering of the magnetic field, while J2 is too close to the core to evaluate its PD.

At optical wavelengths, we also observe emission from a larger volume in the jet than at X-ray energies, with an EVPA within the range of directions of the broad jet. Some of the optical emission also comes from the inner jet region. Because of this, we observe global optical flux enhancement in 2023 August, with respect to 2022 October (with a peak observed in the V filter simultaneous with the X-ray maximum), and optical polarimetric variability.

\begin{figure}[bt]
\centering
\includegraphics[width=0.45\textwidth]{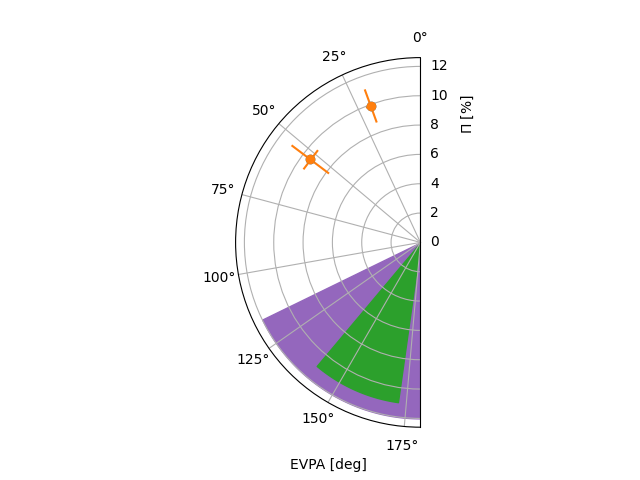}
\caption{X-ray polarization results (orange symbols) and comparison with radio-jet direction: purple area represents the projected radio jet reported in \citealt{Weaver2022}; green area represents the mean radio jet direction from the position angle of knot J1.}
\label{fig:xevpa}
\end{figure}

\begin{figure}[bt]
\centering
\includegraphics[width=0.45\textwidth]{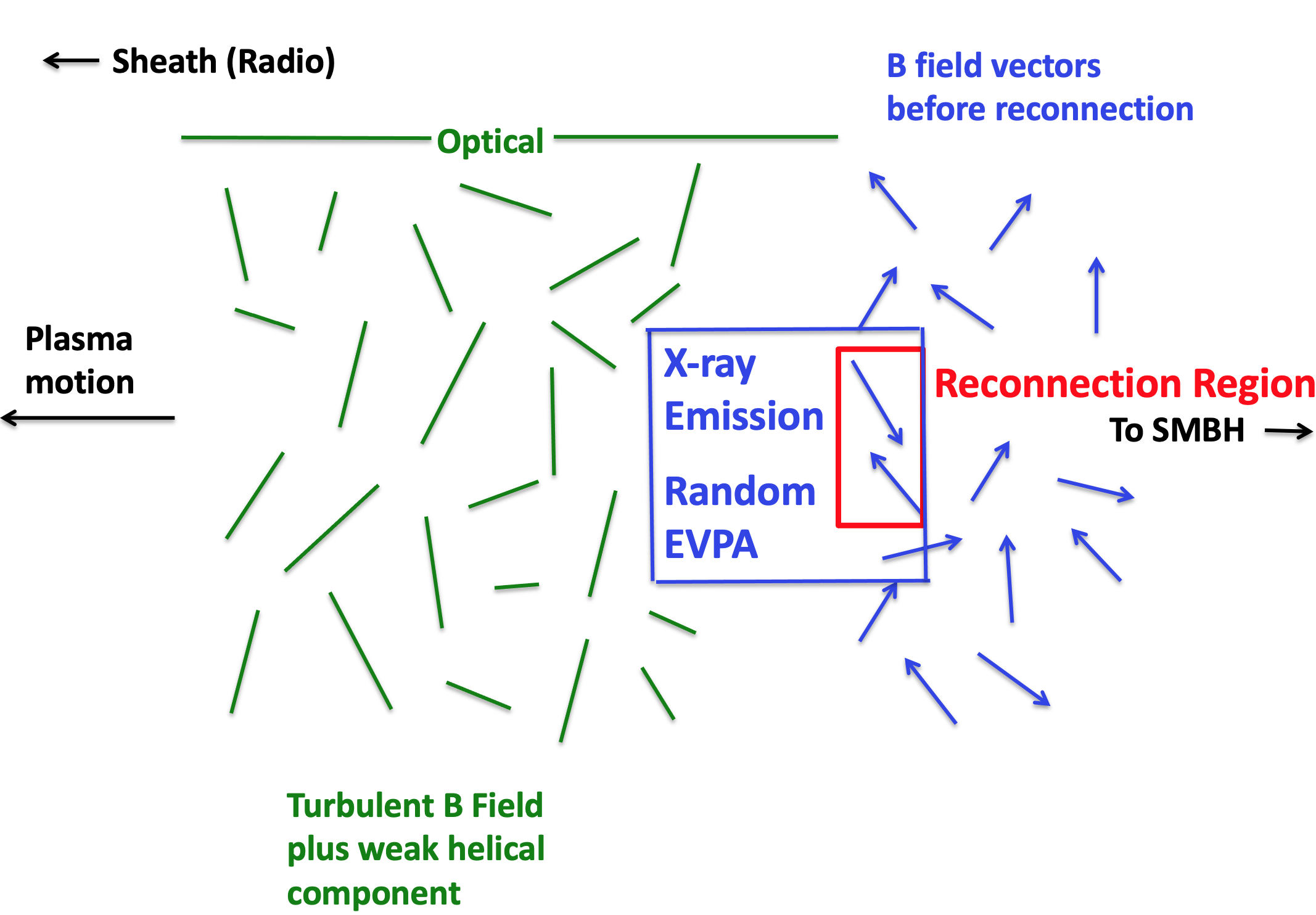}   
\caption{Sketch of possible model for a structured jet in 1ES 1959+650. The X-ray emission arises from a small volume in turbulent plasma where magnetic reconnection occurs, accelerating electrons to extremely high energies. The magnetic field is randomly oriented, with no relation to the jet direction. The optical and radio emission occur downstream, where a helical component of the magnetic field (or weak shocks) provides a net field direction transverse to the jet axis.}
\label{fig:recon_cartoon}
\end{figure}

\begin{figure}[bt]
\centering
\includegraphics[width=0.45\textwidth]{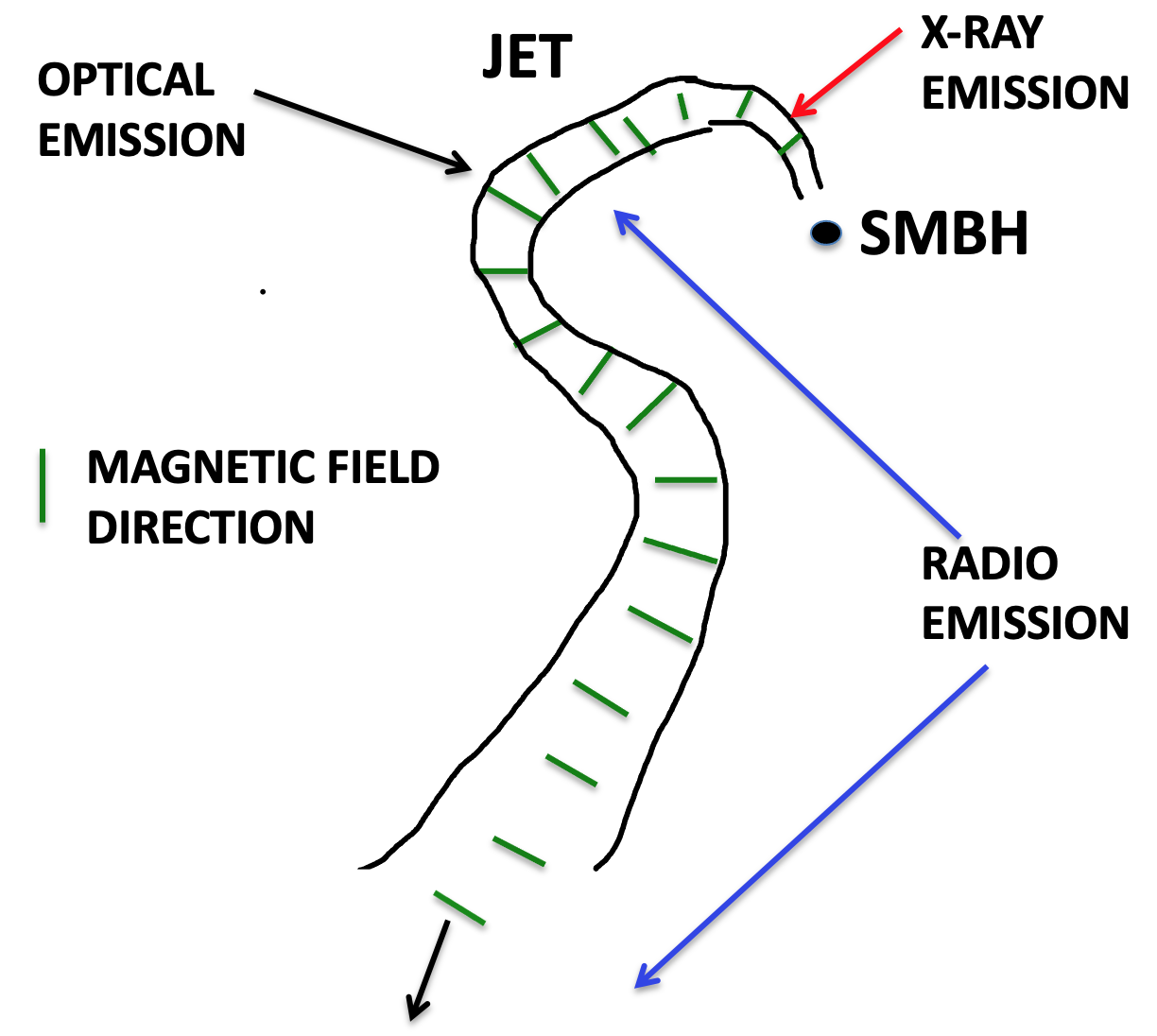} 
\caption{Sketch of a second possible model designed to explain observation of the jet in 1ES~1959+650. The jet is slightly twisted, with bending by less than a few degrees that is accentuated by projection effects. The magnetic field is locally transverse to the jet axis along the jet's length, owing to a weak helical field component or mild shocks. The X-ray emission arises from an upstream region with different EVPA (which is transverse to the field) than that of the lower-frequency emission radiated in the downstream regions.}
\label{fig:bentjet_cartoon}
\end{figure}

\begin{figure}[bt]
\centering
\includegraphics[width=0.42\textwidth]{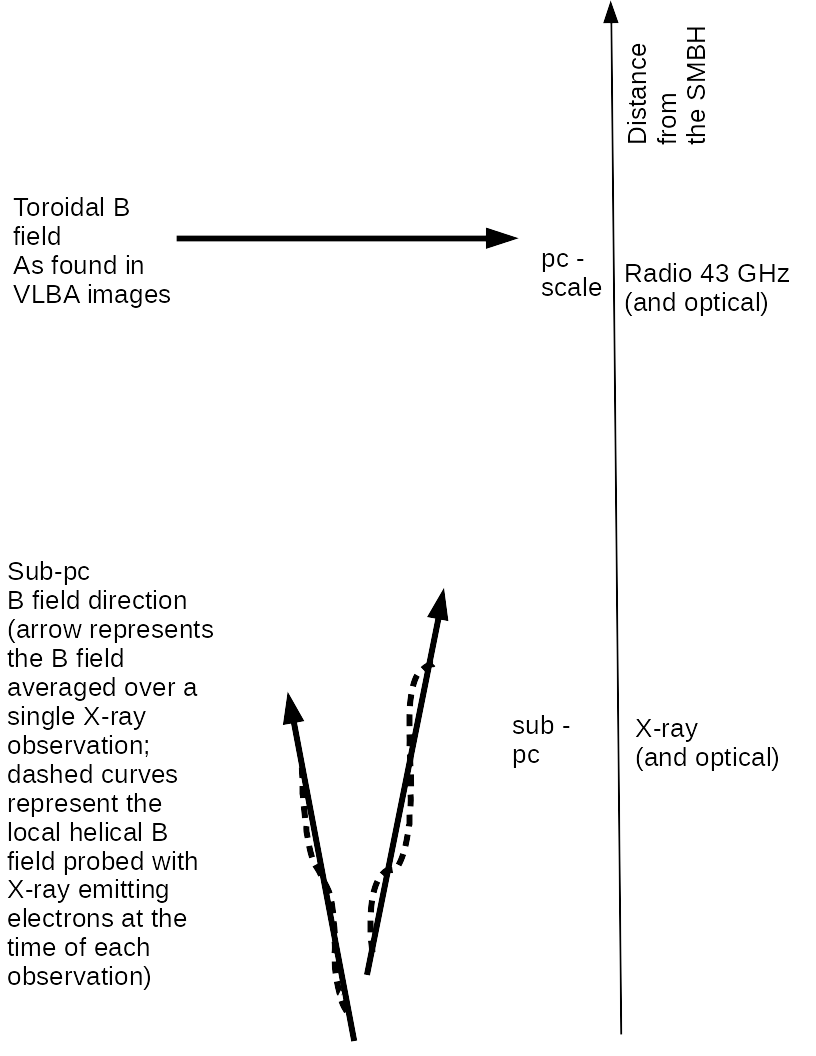}
\caption{Sketch of a third scenario for explaining the X-ray polarization of 1ES~1959+650. X-rays probe sub-parsec scale distances from the SMBH, which contain a helical magnetic field with a small pitch angle (or, alternatively, a turbulent magnetic field with mean direction along the jet axis). Arrows indicate mean direction of the magnetic field during each of our two IXPE observations; X-ray EVPA is perpendicular to the mean magnetic field (projected on the sky). VLBA images exhibit a mainly toroidal magnetic field on parsec scales.}
\label{fig:schema1}
\end{figure}

X-rays are emitted from a more compact region very close to the site of the highest-energy particle acceleration, allowing us to probe the local magnetic field associated with the acceleration process. The X-ray EVPAs are almost orthogonal to the overall jet direction as defined by the VLBA images; see Figure~\ref{fig:xevpa} and \citet{Weaver2022}.
Three rough sketches of possible scenarios are presented in Figures~\ref{fig:recon_cartoon}, \ref{fig:bentjet_cartoon}, and \ref{fig:schema1} \citep[see also][]{Marscher2024}. In Figure~\ref{fig:recon_cartoon}, we assume that plasma is accelerated by local magnetic reconnection in a turbulent zone where oppositely directed magnetic fields stochastically meet. Most X-rays and a small fraction of the observed optical photons are emitted in the magnetic reconnection region, with a polarization angle that is essentially random and unrelated to the jet direction. 
Farther downstream of the accelerating region, the magnetic field includes a component transverse to the jet axis. This could be the result of a weak helical or toroidal field component \citep{Lyutikov2005} or mild compression by shocks \citep{Hughes1985}. Optical photons are also emitted in this region.  The relatively long cooling time of electrons responsible for the optical emission smooths the flaring profile in the optical light curve \citep[see][for further discussion]{tavecchio2021}.

In an alternative scenario, sketched in Figure \ref{fig:bentjet_cartoon}, the jet is twisted by a few degrees, amplified by projection effects owing to the small angle of the axis to the line of sight. The X-ray emission is produced close to the jet origin, with the particle acceleration mechanism unspecified. The jet bends such that the EVPA in the X-ray emitting region happens to lie roughly perpendicular to the downstream jet direction where the 43 GHz core is observed.

The cartoon in Figure~\ref{fig:schema1} represents a scenario that includes a helical magnetic field in the X-ray emission region that could explain the rotating EVPA signals inferred from our analysis above. This could be incorporated into the first scenario (Fig.\ \ref{fig:recon_cartoon}) if there is a small helical field component in the magnetic reconnection region.

%
%

The variability we see in the X-ray EVPA is amplified by the effects of relativistic beaming, aberration, and Doppler shifts. We cannot state from our measurements alone whether the different magnetic field orientations (in X-ray with respect to optical and radio) are due to a spine-sheath structure of the jet (see, e.g., \citealt{Georganopoulos2003,Ghisellini2004,Tavecchio2008}; and
Fig.\ \ref{fig:recon_cartoon}), or if the magnetic field changes continuously from sub-parsec to parsec scales.
In a previous short observation of 50~ks (in 2022 May) IXPE found a low-significance polarimetric signal, and an EVPA almost aligned with the VLBA jet \citep{Errando2024}. IXPE has measured EVPAs lying almost parallel to the jet position angle in other HBLs:  PKS~2155$-$304 \citep{Kouch2024} and Mrk~501 \citep[][after averaging the EVPAs over all 6 IXPE observations]{Chen2024}. The change of X-ray EVPA with respect to the VLBA jet position angle is rather erratic in Mrk~421:
in 2022 May the EVPA was $\sim$51$\degr$ from the VLBA jet axis \citep{2022ApJ...938L...7D}; in 2022 June IXPE observed an EVPA rotation by more than 360$\degr$ over $\sim$5 days \citep{DiGesu2023};
in 2022 December the X-ray EVPA was almost perpendicular to the VLBA jet axis \citep{Kim2023}, and a rotation of the EVPA with a two-component model was detected, so we can apply the scenario depicted in Figure~\ref{fig:schema1};
the IXPE observations in 2023 December revealed an EVPA direction changing from parallel to shifted by $\sim$46$\degr$ with respect to the VLBA jet axis \citep[][ paper submitted]{Maksym2024}.

\section{Conclusions} \label{sec:conclusion}

Based on all of the IXPE observations of blazars, if HBL jets contain a helical magnetic field component on sub-parsec scales, the magnetic field appears not to have a stable configuration. This implies that we can observe different behavior of the X-ray emission in different objects, and at different times in a single object. We can measure long rotations of the EVPA for the case of a magnetic field with symmetry axis along the line of sight (as in the case of Mrk~421 on 2022 June).
If the helical magnetic field has a symmetry axis not perfectly aligned with the line of sight, we can still recognize EVPA rotations by applying the multi-component model; in this case we expect the average X-ray EVPA to be almost orthogonal to the VLBA jet axis. For large off-axis values of the
magnetic field on sub-parsec scales, the EVPA could be parallel to the VLBA jet due to projection effects. We note that radio features in the VLBA images of 1ES~1959+650, PKS~2155$-$304, Mrk~421, and Mrk~501 generally move at subluminal speeds \citep{piner2008,2010ApJ...723.1150P}, implying that the radio jet is aligned within a few degrees of the line of sight.
The turbulent magnetic field scenario on sub-parsec scales could be adopted too, even if it is disfavored as the explanation of the 2022 June IXPE observation of EVPA rotation in Mrk~421. The turbulence model predicts that EVPA rotation could be observed by chance, but the average EVPA should not necessarily be orthogonal to the VLBA jet direction.


The observed average X-ray EVPA of 1ES~1959+650 is transverse to the mean magnetic field direction. We have found variability of the EVPA from one observation to the next, possibly implying that the average magnetic field in the X-ray emitting region is time-variable.

Our results suggest the presence of either a helical magnetic field in the jet of 1ES~1959+650, and perhaps other HBLs, or stochastic processes controlling the direction of the magnetic field within the jet. Further data are needed to confirm this conclusion, define better the temporal distribution of the X-ray EVPA, define better the relationship between the X-ray and longer-wavelength emission regions, and
determine whether stochastic or systematic processes dominate the X-ray EVPA variations.
\newpage
\begin{acknowledgments}
The Imaging X-ray Polarimetry Explorer ({IXPE}) is a joint US and  Italian mission.  The US contribution is supported by the National 
Aeronautics and Space Administration (NASA) and led and managed by its  Marshall Space Flight Center (MSFC), with industry partner Ball 
Aerospace (now, BAE Systems).  The Italian contribution is supported 
by the Italian Space Agency (Agenzia Spaziale Italiana, ASI) through  contract ASI-OHBI-2022-13-I.0, agreements ASI-INAF-2022-19-HH.0 and 
ASI-INFN-2017.13-H0, and its Space Science Data Center (SSDC) with  agreements ASI-INAF-2022-14-HH.0 and ASI-INFN 2021-43-HH.0,
and by the  Istituto Nazionale di Astrofisica (INAF) and the Istituto Nazionale di  Fisica Nucleare (INFN) in Italy.  This research used data products 
provided by the IXPE Team (MSFC, SSDC, INAF, and INFN) and distributed  with additional software tools by the High-Energy Astrophysics Science 
Archive Research Center (HEASARC), at NASA Goddard Space Flight Center  (GSFC). Funding for this work was provided in part by contract 80MSFC17C0012 from the MSFC to MIT in support of the {IXPE} project.  Support for this work was provided in part by the NASA through the Smithsonian Astrophysical Observatory (SAO)
contract SV3-73016 to MIT for support of the {Chandra} X-Ray Center (CXC), which is operated by SAO for and on behalf of NASA under contract NAS8-03060. The IAA-CSIC co-authors acknowledge financial support from the Spanish "Ministerio de Ciencia e Innovaci\'{o}n" (MCIN/AEI/ 10.13039/501100011033) through the Center of Excellence Severo Ochoa award for the Instituto de Astrof\'{i}isica de Andaluc\'{i}a-CSIC (CEX2021-001131-S), and through grants PID2019-107847RB-C44 and PID2022-139117NB-C44. 
Some of the data are based on observations collected at the Observatorio de Sierra Nevada, owned and operated by the Instituto de Astrof\'{i}sica de Andaluc\'{i}a (IAA-CSIC). Further data are based on observations collected at the Centro Astron\'{o}mico Hispano-Alem\'{a}n(CAHA), operated jointly by Junta de Andaluc\'{i}a and Consejo Superior de Investigaciones Cient\'{i}ficas (IAA-CSIC). The Submillimetre Array is a joint project between the SAO and the Academia Sinica Institute of Astronomy and Astrophysics and is funded by the Smithsonian Institution and the Academia Sinica. Mauna Kea, the location of the SMA, is a culturally important site for the indigenous Hawaiian people; we are privileged to study the cosmos from its summit.
\end{acknowledgments} 
\begin{acknowledgments} 
The data in this study include observations made with the Nordic Optical Telescope, owned in collaboration by the University of Turku and Aarhus University, and operated jointly by Aarhus University, the University of Turku and the University of Oslo, representing Denmark, Finland and Norway, the University of Iceland and Stockholm University at the Observatorio del Roque de los Muchachos, La Palma, Spain, of the Instituto de Astrofisica de Canarias. 
The data presented here were obtained in part with ALFOSC, which is provided by the Instituto de Astrof\'{i}sica de Andaluc\'{i}a (IAA) under a joint agreement with the University of Copenhagen and NOT. E.\ L.\ was supported by Academy of Finland projects 317636 and 320045. We acknowledge funding to support our NOT observations from the Finnish Centre for Astronomy with ESO (FINCA), University of Turku, Finland (Academy of Finland grant nr 306531). The research at Boston University was supported in part by National Science Foundation grant AST-2108622, NASA NuSTAR Guest Investigator subcontract through JPL (RSA no. 1691953), 
NASA {Fermi} Guest Investigator grants 80NSSC23K1507 and 80NSSC23K1508, and NASA {Swift} Guest Investigator grant 80NSSC22K0537. This study used observations conducted with the 1.8 m Perkins Telescope Observatory (PTO) in Arizona (USA), which is owned and operated by Boston University. The above study is based in part on observations obtained  with 
{XMM-Newton}, an ESA science mission with instruments and contributions directly funded by ESA Member States and NASA. This research has made use of data from the RoboPol programme, a collaboration between Caltech, the University of Crete, IA-FORTH, IUCAA, the MPIfR, and the Nicolaus Copernicus University, which was conducted at Skinakas Observatory in Crete, Greece. I. L. was supported by the NASA Postdoctoral Program at the Marshall Space Flight Center, administered by Oak Ridge Associated Universities under contract with NASA. Partly based on observations with the 100-m telescope of the MPIfR (Max-Planck-Institut f\"ur Radioastronomie) at Effelsberg. Observations with the 100-m radio telescope at Effelsberg have received funding from the European Union’s Horizon 2020 research and innovation programme under grant agreement No 101004719 (ORP). The Very Long Baseline Array is an instrument of the National Radio Astronomy Observatory. The National Radio Astronomy Observatory is a facility of the National Science Foundation operated by Associated Universities, Inc.
\end{acknowledgments}
\clearpage 

\facilities{Calar Alto, Effelsberg 100m  radio telescope, IXPE, LX-200,  NuSTAR, XMM Newton, NOT, Perkins, Skinakas observatory, SMA, Observatorio de Sierra Nevada (SNO), Swift(XRT and UVOT)}

\appendix

\section{likelihood estimator for the two-component model} 
\label{appendix:cash}
The two-component model assumes that the polarized emission is obtained with two independent components: the first (steady) component with fixed PD $\Pi_1$ and angle $\Psi_1$ (EVPA does not rotate), and a second (rotating) component characterized by a rotating EVPA with constant rotation velocity $\omega_2$, and fixed PD $\Pi_2$; the polarization angle of the rotating component at the beginning of observation is denoted by $\Psi_2(t=0)$. 
In this simple model, parameters do not depend on energy. 
If the fluxes of the two components are $F_1$ and $F_2$, respectively, we define the relative fluxes of the two components as $R_1=F_1/(F_1+F_2)$ and  $R_2=F_2/(F_1+F_2)$.
We can parametrize the two components with Stokes parameters $q_1$, $u_1$, $q_2(t)$, $u_2(t)$ (with Stokes parameters for the second component that vary with time) or $q_1$, $u_1$, $q^0_2$, $u^0_2$
of the bi components, and with the angular velocity $\omega$ of the EVPA of the second component (where $q^0_2$ and $u^0_2$ are the Stokes parameters of the second component at t=0). We can generalize the event density proposed for the likelihood estimator in Equation~(51) of \cite{2021ApJ...907...82M}: The event density for the case of the two-component model should not be integrated over the time interval, because it changes with time owing to the polarization rotation. The event density in this case is \\
\begin{multline}
\lambda(E,\psi)= 
\big[R_1\big(1+\mu(E) q_1\;\! \cos(2\psi)+\mu(E) u_1\;\! \sin(2\psi)\big)+
R_2\big(1+\mu(E) q_2(t)\;\! \cos(2\psi)+\mu(E) u_2(t)\;\! \sin(2\psi)\big)\big]f_E A_E\;\! {\rm d}t\;\! {\rm d}E \;\! 
 {\rm d}\psi \ , 
\label{eq:mainevdensity}
\end{multline}
where we use the same notation adopted in \cite{2021ApJ...907...82M}.
The Stokes parameters $\big(q_2,u_2\big)$ describe a rotation of the EVPA, and can be written: $q_2(t)=\Pi_2 \cos\big(2(\Psi^0_2+\omega t)\big)$ and $u_2(t)=\Pi_2\sin\big(2(\Psi^0_2+\omega t)\big)$. Using the
Werner formulas twice, performing the same steps used in \cite{2021ApJ...907...82M} to obtain their eq. 54 from eq. 51,
and in the assumption of no energy dependence of polarization, we can obtain  an expression for the likelihood estimator of the two-component model: 
\begin{multline}
-2\sum_i \ln \big[ 1+  \mu_i R_1q_1\;\! \cos(2\psi_i)+\mu_i R_1 u_1\;\! \sin(2\psi_i)+\mu_i (1-R_1)q^0_2\;\! \cos(2(\psi_i-\omega t_i))+\mu_i (1-R_1)u^0_2\;\! \sin(2(\psi_i-\omega t_i)) \big] \ .
\end{multline}
Here, $u^0_2$, $q^0_2$ are the Stokes parameters for the rotating component evaluated at $t=0$.
There are 6 physical parameters to describe the model ($R_1$, $u_1$, $q_1$, $u^0_2$, $q^0_2$, $\omega$), but model fitting requires only 5 independent parameters, e.g.: $R_1q_1$, $R_1u_1$, $(1-R_1)q^0_2$, $(1-R_1)u^0_2$, and $\omega$.
With a few steps we obtain an expression involving  PD and EVPA of the bi components:\\
\begin{multline}
S(R_1,\Pi_1,\Psi_1,\Pi_2,\Psi_2,\omega)=
-2\sum_i \ln \biggr[
1+\mu_iR_1\Pi_1\biggr(\cos(2\Psi_1)\cos(2\psi_i)+\sin(2\Psi_1)\sin(2\psi_i)\biggr)  \\ 
+ \mu_i(1-R_1)\Pi_2\biggr(\cos(2\Psi^0_2)\cos(2(\psi_i-\omega t_i)+\sin(2\Psi^0_2)\sin(2(\psi_i-\omega t_i  ) \biggr)\biggr] \ .
\end{multline}      
The last expression is the likelihood estimator used in this study.\\ 

We can rewrite Equation~(\ref{eq:mainevdensity}) as\\
\begin{equation}
\lambda(E,\psi)\ =\  \big[1+\mu(E)q_T(t)\cos(2\psi)\ +\ \mu(E) u_T(t)\sin(2\psi)\big]f_E\;\! A_E\;\! {\rm d}t\;\!{\rm d}E \;\! {\rm d}\psi \ ,
\label{eq:vect_evdensity}
\end{equation}
where $q_T(t)=R_1q_1+R_2q_2(t)$ and $u_T(t)=R_1u_1+R_2u_2(t)$.
Equation~(\ref{eq:vect_evdensity}) shows
the known Stokes summation rule for incoherent polarized components: at any given time we can represent
the two-component model in the $(q,u)$ plane as a vector $(q_T(t),u_T(t))$ that is the weighted sum of $(q_1,u_1)$ and $(q_2(t),u_2(t))$
with weights given by the relative fluxes $R_1$ and $R_2$ of the intervening components.
For any given observation with a given exposure, the two-component model is represented in the $(q,u)$ plane
as a circle of radius $R_2\Pi_2$, and with offset $R_1\Pi_1$ from the origin of the coordinates. In general, the event density of 
a multi-component  model can be represented at any given time as the weighted sum of $(q_i,u_i)$ with weights
$R_i$ corresponding to the relative flux of the $i$ components:\\
\begin{equation}
\left( \begin{array}{c}
q_T\\
u_T
\end{array} \right)
\ =\ \sum_i R_i
\left( \begin{array}{c}
q_i\\
u_i
\end{array}  \right) \ . 
\label{eq:vect_repres}
\end{equation}
\section{Adding a counter-rotating component to the two-component model (three-component model with two counter-rotating components)} 
\label{appendix:threecomp}
Following the vectorial representation in Equation~(\ref{eq:vect_repres}), we can evaluate
the effect of a counter-rotating component in the $(q,u)$ plane (with the counter-rotating component having opposite rotation velocity with respect to the rotating component). In the simple case of a rotating
and counter-rotating component with the same phase at $t=0$, the two components add together coherently
because they have the same direction.
After a quarter of a turn in the $(q,u)$ plane the two components have opposite directions,
reducing the total PD.
After a half turn in the $(q,u)$ plane, the two components add coherently, but with an opposite direction
with respect to the case at $t=0$.
After three-quarters of a turn, the situation is analogous to the case at a quarter of a turn, but with opposite
direction. In general, a model with three components (a constant polarization component and both rotating and counter-rotating components) draws an ellipse in the $(q,u)$ plane. The orientation of the axes of the ellipse depends on the relative phase of the rotating and counter-rotating components. In the special case of oppositely rotating components with the same relative fluxes, the ellipse reduces to a segment in the $(q,u)$ plane; and the polarization vector oscillates within the segment.

The event density for this model (the three-component model with two counter-rotating components) can be evaluated
directly using the formalism of Equation~(\ref{eq:vect_repres}), within which the event density can be written as
\begin{equation}
\left( \begin{array}{c}
q_T\\
u_T
\end{array} \right)
\ =\ R_1
\left( \begin{array}{c}
q_1\\
u_1
\end{array}  \right) \ 
\ +\ R_2
\left( \begin{array}{c}
q_2(t)\\
u_2(t)
\end{array}  \right) \ 
\ +\ R_2^*
\left( \begin{array}{c}
q_2^*(t)\\
u_2^*(t)
\end{array}  \right) \ . 
\label{eq:vect_repres_threecomp}
\end{equation}
We can write the vectors ($q_i,u_i$) as the product of the PD of the $i$ component and a unit vector $\hat{p}_i$ containing the information of the EVPA for the $i$ component: $(q_i,u_i)=\Pi_i\hat{p}_i$. Expression (\ref{eq:vect_repres_threecomp}) can be rewritten as  
\begin{align}
\left( \begin{array}{c}
q_T\\
u_T
\end{array} \right)
& = R_1\Pi_1\hat{p}_1  +  R_2\Pi_2\hat{p}_2 \ +\  R_2^*\Pi_2^*\hat{p}_2^*  \nonumber \\ 
& = R_1\Pi_1\hat{p}_1  +  
(R_2\Pi_2+R_2^*\Pi_2^*)  
\Bigg[ 
\frac{R_2\Pi_2}{R_2\Pi_2+R_2^*\Pi_2^*}\hat{p}_2  
 +  \left(1-\frac{R_2\Pi_2}{R_2\Pi_2+R_2^*\Pi_2^*}\right)\hat{p}_2^*  
\Bigg] \ . 
\label{eq:vers_repres_threecomp}
\end{align}

\section{Details of the Analysis of X-ray data for the 2022 October observation}
\label{appendix:signifA}
In order to check the result of the unbinned log-likelihood analysis, we applied the binned analysis and the $\chi^2$ statistics for both the constant model and the two-component model.
The results are reported in Figure~\ref{fig:const_vs_bicomp_oct2022}. While the binned analysis is less sensitive, we determined that the difference of $\chi^2$  for the two nested models is close to the difference of the log-likelihood obtained with the unbinned analysis. Therefore, our results are validated.

We used Monte-Carlo simulations to investigate the statistical distribution of $\Delta C$, where $C$ is the log-likelihood minimum, in the case of sets of data simulated with constant polarization:
The two-component model has 5 parameters (including a frequency), while the constant polarization model has two parameters.
With the  candidate frequency of the rotating component fixed, $\Delta C$ for the two-component model relative to the constant
polarization model has a $\chi^2$ distribution with 2 degrees of freedom.
We also verified the $\Delta C$ distribution on real data: we analysed all the IXPE observations of point-like, non-blazar sources with a count rate $<$ 25~count\,s$^{-1}$ performed within the first two years of observations. Observations lasting more than 6 days were subdivided into slots at least 3 days long. We obtained a total of 
142 slots.
We tried frequencies from the base frequency (${0.5}/{T_{\rm exposure}}$, where the factor 0.5 comes from the fact that the EVPA is confined within a 180$\degr$ range) up to 10~turn~d$^{-1}$. With the exclusion of Cyg~X-3, we obtained that the $\Delta C$ distribution follows a $\chi^2$ distribution when computed as a function of EVPA rotating frequency. For Cyg~X-3 we were able to reconstruct the orbital frequency from the frequency scan. Therefore, we consider the statistics and the procedure to be validated with both simulations and real data. Moreover, with an ad-hoc scan we were able to reconstruct the spinning frequency of the X-ray pulsar GX~301$-$2. We also applied the method to the IXPE observations of Mrk~421 on June 2022. We obtained the same findings reported in \cite{DiGesu2023}.
Finally, the method has been already used to find the EVPA rotation during the IXPE observation of Mrk~421 on 2022 December; and the result has been validated with $\chi^2$ fit of X-ray polarimetric evolution of the source \citep{Kim2023}.\\

We estimate the number of trials from 
the frequency range we searched over (approximately 10 times the  base frequency. 
While performing the frequency scan to minimize the log-likelihood estimator, we have to deal with random signals. Random signals in frequency domain have a correlation width that corresponds to the base frequency, so the number of
independent trials is the ratio of the frequency range divided by the frequency width of random signals.        
The scan on frequencies was performed for both positive and negative frequencies, hence the frequency range should be multiplied by a factor of two. We then have 20 independent trial frequencies.
The chance probability can be obtained by using the binomial distribution for
at least a signal with $\Delta C \ge 19.6$ and 20 trials. We then derive a
probability of $1.1\times 10^{-3}$ that the two-component model provides a
better fit to the data (with respect to the constant polarization model) by
random chance.

\section{Details of the Analysis of X-ray data for the 2023 August observation}
\label{appendix:signifB}
We studied with Monte-Carlo simulations the case of applying the three-component model to event lists simulated with a constant EVPA and PD.
The probability $P(\Delta C^*,N)$ to find by chance at least one signal with $\Delta C^*$ from a sample of $N$ extractions depends on the probability $p^*$ of extracting a value of $\Delta C^*$ in a single trial
(which can be computed from the $\chi^2$ distribution with 4 degrees of freedom):
$P(\Delta C^*,N)=\sum_{k=1}^{N}B(k,p^*)$,
where $B(k,p)$ is the binomial distribution for $k$ successes, and $p$ is the probability of having a success in a single trial.
With the chosen window length ($\Delta t$), random signals have a frequency width of 
${0.5}/{\Delta t}$ in the frequency scan reported in Figure \ref{fig:dcash2023};  we have 42 independent trials per window.
For a single window we have 
$P(\Delta C=26.6,N=42)\ =\ 1.0\times 10^{-3} $, and 
$P(\Delta C=20.5,N=42)\ =\ 1.6\;\!\%$. If we consider that the search was performed on four staggered windows, the number of independent trials should be multiplied by a factor of $\le$4.
Therefore, the chance probabilities are $\le$ 4.0$\times 10^{-3}$ and 6.5\;\!\% for the candidate frequencies at 5.2$\pm$0.1 and 1.9$\pm$0.1 turn~d$^{-1}$, respectively.

\begin{figure*}[tb]
\centering
\begin{tabular}{lll}
\includegraphics[width=0.32\textwidth]{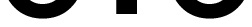} &
\includegraphics[width=0.32\textwidth]{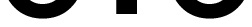} &
\includegraphics[width=0.32\textwidth]{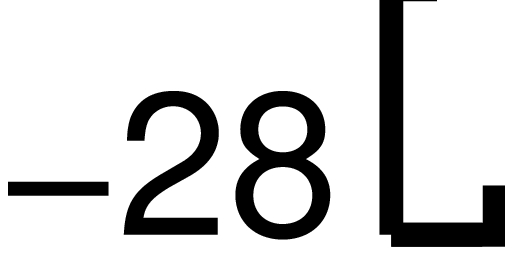} \\
\end{tabular}
\caption{Results of binned analysis fitting of the 2023 August polarimetric light curve. Fit was performed on the $q$ and $u$ Stokes light curves; Left panel: Reduced $\chi^2$ for the constant polarization model as a function of the number of time bins, and associated probability that the fit function is true. Central panel: Reduced $\chi^2$ for the three-component polarization model as a function of the number of time bins, and associated probability that the fit function is true. Right panel: difference between the $\chi^2$ for the three-component model and the $\chi^2$ for the constant component as a function of the number of time bins.}
\label{fig:const_vs_tricomp_aug2023_w3}
\end{figure*}

We analysed the folded polarimetric light curves reported in Figure~\ref{fig:phased_pol_lc} with a binned $\chi^2$ method to validate the unbinned analysis for window 3.  We tested the two nested hypotheses of the constant and three-component models. Results are shown in Figure~\ref{fig:const_vs_tricomp_aug2023_w3} as a function of the number of time bins. The $\Delta \chi^2$ between the constant and three-component models is $\Delta \chi^2\ \sim25$, confirming the unbinned likelihood result.\\

We tested with simulations whether the gaps in the observation could produce a spurious rotating polarization signal,
but we were unable to reproduce any such signal.
We also investigated whether the 
satellite dithering strategy could result in some characteristic frequency. In fact, noisy readout pixels, or charge build-up in the GEM \citep{BALDINI2021102628}, combined with satellite
dithering, could generate spurious, time variable polarization.
Moreover, a spurious polarization measurement can be obtained for a source with projected image close to the detector boundaries \citep{Di_Marco_2023}.
We analyzed attitude data of the IXPE satellite, and studied the angular displacements of the source position (in detector D1 frame) with respect
to the x and y axis of the D1 frame, along the radial direction, and rotations about the normal axis.
We found several typical frequencies: the lower frequencies are at 14.9\;d$^{-1}$ (the satellite orbital frequency) and at  64.9\;d$^{-1}$,
with similar strength (we also found  secondary harmonics and a beat frequency at 
50\;d$^{-1}$),
incompatible with our results.
 In order to verify the effect on polarization results, we executed a further
 analysis of our window 3 data subset, extending the frequency scan over the
 first main frequency. We found only a weak signal, with $\Delta C=-13.7$
 peaking at 14.84\;d$^{-1}$, that could be tentatively correlated with the
 dithering strategy. This weak signal obtained for window 3 is not confirmed
 over the entire integration, for which we should expect a cumulative
 amplitude. Hence, we conclude that the combined effect of dithering and
 charge buildup have negligible effect on polarization rotation for a source
 photon count rate of $\sim2$ cts~s$^{-1}$.

\bibliography{ixpe_obs_1ES1959p650.bib}
\bibliographystyle{aasjournal}


\end{document}